\documentclass[aps,prb,twocolumn,groupaddress,floatfix]{revtex4}

\usepackage{graphicx}

\usepackage{amssymb}

\usepackage{amsmath}

\begin{document}

\title{Shape memory ferromagnets}

\author{A.~N.~Vasil'ev}
\affiliation{Physics Department, Moscow State University, Moscow
119992, Russia}

\author{V.~D.~Buchel'nikov}
\affiliation{Chelyabinsk State University, Chelyabinsk 454021,
Russia}

\author{T.~Takagi}
\affiliation{Institute of Fluid Science, Tohoku University, Sendai
980-8577, Japan}

\author{V.~V.~Khovailo}
\affiliation{National Institute of Advanced Industrial Science and
Technology, Tohoku Center, Sendai 983-8551, Japan}

\author{E.~I.~Estrin}
\affiliation{I.P.~Bardin Central Research Institute of Ferrous
Metallurgy, Moscow 107005, Russia}

\begin{abstract}
In ferromagnetic alloys with shape memory large reversible strains
can be obtained by rearranging the martensitic domain structure by
a magnetic field. Magnetization through displacement of domain
walls is possible in the presence of high magnetocrystalline
anisotropy, when martensitic structure rearrangement is
energetically favorable compared to the reorientation of magnetic
moments. In ferromagnetic Heusler alloys Ni$_{2+x}$Mn$_{1-x}$Ga
the Curie temperature exceeds the martensitic transformation
temperature. The fact that these two temperatures are close to
room temperature offers the possibility of magnetically
controlling the shape and size of ferromagnets in the martensitic
state. In Ni$_{2+x}$Mn$_{1-x}$Ga single crystals, a reversible
strain of $\sim 6$\% is obtained in fields of $\sim 1$~T.
\end{abstract}

\maketitle

\vspace{5cm}

\bigskip

\section{Introduction}

In this review we examine the properties of ferromagnets that
exhibit shape memory and hyperelasticity (superplasticity) and
that make it possible to control these effects magnetically. Shape
memory in such substances is related to a martensitic phase
transition, and the effect of a magnetic field on the parameters
of the martensitic phase is caused by magnetoelastic interaction.
The specific feature of the magnetoelastic interaction is that the
entities that participate in the interaction are large
intercorrelated ensembles of structural and ferromagnetic domains.
In this respect the magnetoelastic interaction in shape memory
ferromagnets, which leads to giant magnetically induced strains,
differs significantly from magnetostriction, which is essentially
a one-particle effect.

The possibility that certain substances can be reversibly
controlled both in shape and size by mechanical stresses and
electric or magnetic field places them into a separate class of
functional materials. Representatives of this class are
magnetostrictive materials, piezoelectric materials, and materials
with shape memory. The hierarchy of the attainable reversible
strains in crystals caused by magnetostriction (MS), the
piezoelectric effect (PE), and shape memory (SM) is as follows:

$$ \frac{\Delta L_{\mathrm{MS}}}{L} \sim 10^{-3}, \frac{\Delta
L_{\mathrm{PE}}}{L} \sim 10^{-2}, \frac{\Delta L_{\mathrm{SM}}}{L}
\sim 10^{-1}.$$

Each of these methods of controlling the size of samples of such
materials has its advantages and drawbacks and its own area of
application. Magnetostrictors are used as emitters and detectors
of sound, frequency stabilizers, and delay lines in wireless and
acoustic devices, as microactuators and as magnetomechanical
relays. Materials with shape memory are used in thermosensitive
force-summing elements, detachable and nondetachable joints that
do not require welding or soldering, and in various types of
clamps used in medicine. An advantage of magnetostrictive
materials is the small response time, while an advantage of
materials with shape memory is the large reversible strains.

Studies of shape memory ferromagnets focus on ways to produce
materials in which short response times are combined with large
reversible strains. Several Heusler alloys and the intermetallic
compounds Co--Ni, Fe--Pd, Fe--Pt, and Fe--Ni--Co--Ti have been
investigated in this connection. The most promising results have
been obtained with the ferromagnetic Heusler alloy Ni$_2$MnGa.
This unique alloy (more exactly, the family of
Ni$_{2+x+y}$Mn$_{1-x}$Ga$_{1-y}$ alloys) has made it possible to
achieve magnetically controlled variations in the linear size of
crystals up to 6\%, which is the theoretical limit of strains for
martensitic transformations in this material.

The general ideas about martensitic transformations and the
related special features of the mechanical behavior of alloys with
thermoelastic martensitic behavior are examined in Section~2. In
Section~3 we discuss the crystal structure of the Heusler alloys
Ni$_{2+x+y}$Mn$_{1-x}$Ga$_{1-y}$ and in Section~4, the magnetic
properties of these alloys. Section~5 is devoted to a description
of the main physical properties of
Ni$_{2+x+y}$Mn$_{1-x}$Ga$_{1-y}$ as a function of composition. In
Section~6 we describe experimental implementations of the
magnetically controlled shape-memory effect in
Ni$_{2+x+y}$Mn$_{1-x}$Ga$_{1-y}$. A summary of the data on other
materials for which magnetic control of the shape memory effect is
possible is given in Section~7. Section~8 is devoted to the theory
of structural and magnetic phase transitions in cubic
ferromagnets. In Section 9 we discuss the prospects of further
investigations in this field and the possibilities of using this
class of materials in applications.

\section{Martensitic transformations
and the shape memory effect}

\subsection{Thermoelastic and nonthermoelastic martensitic
transformations}

Martensitic transformations are structural phase transitions of
the diffusionless, cooperative type. The characteristic features
of such transformations are the cooperative displacements of
neighboring atoms by distances smaller than the atomic separation,
a rigorous crystallographic connection between the lattices of the
initial and final phases, and changes in the shape of the
transformed region [1].

Martensitic transformations were first discovered in iron-based
alloys (steel). Initially they were interpreted as structural
transformations of a high-temperature face-centered cubic (fcc)
phase ($\gamma$-phase, austenite) into a low-temperature
body-centered cubic (bcc) phase ($\alpha$-phase, martensite). The
main laws governing such transformations have been established
without doubt. It was found that transformations similar to the
martensitic transformation in steel occur in solids of a different
nature (metals, insulators, semiconductors, and organic compounds)
and belong to one of the main types of phase transformations in
the solid state [2--4].

The most general feature of martensitic transformations is that
they occur in the solid medium at low temperatures, when all
diffusion processes are frozen. The conditions under which
martensitic transformations take place (elastic medium, low
temperatures) determine all the main features of the
transformations. The manifestations of martensitic transformations
are multivarious. They may proceed athermally with a rate equal to
the speed of sound, or they can be thermally activated and develop
with a measurable rate, be reversible or irreversible, or lead to
formation of morphologically different structures that depend on
the crystal geometry and properties of the initial and final
phases and on the development of relaxation processes. Among
transformations that are called martensitic are those that may be
considered "nearly second-order" and transformations that are
clearly first-order accompanied by large heat and volume effects
and a sizable hysteresis between the direct and reverse
transformations.

The emergence of a new-phase crystal inside the initial matrix
leads to the appearance of elastic and surface energies. If the
difference in the crystal lattices is slight and the accommodation
of the new-phase and matrix crystals is not accompanied by an
irreversible plastic strain, the transformation may be considered
fully reversible. The hysteresis between the direct and reverse
transformations in this case is small, so that the structure of
the initial phase is completely restored as a result of the
reverse transformation. The small elasticity modulus and the high
elastic limit of the phases (which ensures elastic accommodation
of the intrinsic transformation strain), in addition to the small
difference in the lattices of the initial and final phases, are
conducive to the reversible nature of the transformation.
Martensitic transformations of this type are called thermoelastic.
The reversible nature of the transformation is the necessary
condition for emergence of reversible effects under an external
load: hyperelasticity, rubber-like behavior, and the shape memory
effect.

If the difference in the crystal lattices of the phases is large
and cannot be elastically accommodated, the transformation is
accompanied by plastic strain and the emergence of structural
defects, which hinder the easy motion of interphase boundaries.
Here the reverse transformation proceeds not so much because of
the gradual decrease in the size of the martensite crystals but
largely because of nucleation and the growth of austenite crystals
inside the martensitic matrix. This process is accompanied by an
increase in the number of orientations of the high-temperature
phase, in contrast to the reproduction of the initial orientation
in a thermoelastic transformation. Martensitic transformations of
this type are called nonthermoelastic.

Since martensitic transformations are first-order phase
transitions, the temperatures of their beginning and end are their
characteristic parameters. The transformation of austenite into
martensite (direct transformation) is characterized by the
temperature $M_s$ of the appearance of nuclei of the martensitic
phase in the austenitic matrix and the temperature $M_f$ at which
the formation of the martensite ends. In the reverse
transformations these temperatures are denoted $A_s$ and $A_f$.

There is no well-defined boundary between thermoelastic and
nonthermoelastic transformations: to one extent or another all
martensitic transformations are reversible --- the question is how
large the size of the hysteresis between the direct and reverse
transformations is. In some cases (with Cu--Sn, Fe--Pt, and
Fe--Co--Ni--Ti alloys) thermal treatment can change the size of
the hysteresis and thus change the type of martensitic
transformation.

Anomalies in the mechanical properties are inherent in all
low-temperature structural phase transformations, irrespective of
the nature of the material and the electrical and magnetic
properties of the phases. Specific effects related to the action
of an electric or magnetic field manifest themselves only when one
of the phases exhibits ferroelectric or ferromagnetic properties,
respectively.

If, due to the crystal geometry, the martensitic transformation is
of the "nearly second-order" type and the intrinsic transformation
strain $\varepsilon$ is the order parameter, the elastic constant
corresponding to this strain vanishes at the start of the
transformation or becomes small. Examples are the low-temperature
structural transitions in the superconductors Nb$_3$Sn and
V$_3$Si~[5]. In substances in which the crystal geometry of the
transformations is such that they could, in principle, be
second-order transitions but proceed as first-order
transformations (transformations in the In--Tl, Au--Cd, and Mn--Cu
alloys), the elastic constants decrease as the point at which the
transformation begins is approached but remain finite~[5, 6].
Finally, if the transformation is clearly a first-order one, as it
happens to be in Li, Na, and Cs [7] or in the transition of the
high-temperature face-centered phase to the low-temperature
body-centered phase in iron alloys, the elastic constants have no
anomalies as the martensite point is approached.

\subsection{Hyperelasticity and superplasticity}

Above the martensite point a transformation can be caused by
applying external stresses~[8]. An external stress performs work
along the path determined by the intrinsic transformation strain
$\varepsilon$. This work provides an additional contribution to
the thermodynamic driving force of the transformation. The shift
in the temperature $T_0$ of the thermodynamic equilibrium of the
phases and, respectively, in the martensite point $M_s$ is
described by the generalized Clapeyron equation

$$\frac{dT_0}{d\sigma _{ij}} = \frac{\varepsilon}{\Delta S}$$

\noindent where $\sigma _{ij}$ are the external stresses
corresponding to $\varepsilon$ and $\Delta S$ is the change in
entropy at the transformation. For fixed $\varepsilon$ and $\Delta
S$, the shift of the martensite point is greater, the higher the
external stresses. However, under stresses greater than the
elastic limit plastic strain sets in, and this hinders
thermoelastic transformation. This limits both the level of
admissible stresses and the greatest possible rise in the
martensite point at which the transformation remains
thermoelastic. If the temperature $M_{\sigma}$ (the temperature at
which the transformation can be caused by stresses not exceeding
the elastic limit) is higher than the temperature $A_f$ at which
the reversible transformation ends, when the external stresses are
lifted the martensite that formed under the stresses transforms
into the initial phase, i.e., the transformation proves to be
mechanoelastic. Thus, within a certain temperature range that
adjoins the martensite point, under an external load there occur a
mechanoelastic martensitic transformation and a reversible
"hyperelastic" strain associated with this transformation.

In some alloys subjected to external stresses there can be one or
several martensitic transformations of one martensitic phase into
other martensitic phases. In this case hyperelasticity caused by
these mechanoelastic transformations may be observed even below
the martensite point.

In some alloys subjected to an external load in the temperature
interval from $M_s$ to $A_s$, the martensite that was produced as
a result of external stresses does not disappear when the load is
removed and, instead of hyperelasticity, there is residual strain
caused by the emergence of martensite (superplasticity). At $M_s$
the stresses that initiated the transformation and brought on
pseudoplastic strain are close to zero. As the temperature is
raised, the stresses that caused superplastic strain increase,
too, in accordance with the generalized Clapeyron equation. At a
certain "distance" from the martensite point the stresses that
initiated the transformation may exceed the yield stress of the
material and ordinary plastic strain sets in along with the
transformation. Under certain conditions the interaction of
plastic strain and martensite transformation may lead to a
substantial increase in plastic strain, which precedes the breakup
of the sample.

\subsection{Shape memory effect}

In most alloys with a thermoelastic martensite transformation the
application of a load in the martensitic phase leads to residual
strain in the material. This strain sets in because of the
transformations in the martensitic structure (via twinning, growth
of martensite crystals favorably oriented with respect to the
external load at the expense of less favorably oriented crystals,
etc.). As the temperature is increased, the initial orientation
and structure of the high temperature phase, and thus the initial
shape of the sample, are restored in the reverse transformation.
The effect of restoration of the shape of the deformed sample as a
result of a reverse martensitic transformation in the process of
heating the sample is called the shape memory effect.

Restoration of shape in the reverse transformation is observed in
thermoelastic martensitic transformations as well as in
nonthermoelastic martensitic transformations. A specific feature
of alloys with thermoelastic transformation is that the degree of
shape restoration in them is very high and approaches 100\%. All
factors that favor the thermoelastic nature of a martensitic
transformation also favor the maximum manifestation of the shape
memory effect.

Since the formation of nuclei of the low-temperature phase at a
martensitic transformation is heterogeneous, a fact corroborated
by the microstructural reversibility of the transformation, the
creation of the most favorable locations of nucleation of
martensite can determine the entire sequence of the emergence of
martensite crystals and, consequently, the overall change of shape
of the sample in the transformation. The creation and pinning of
the centers of heterogeneous nucleation makes it possible to
control the process of martensitic transformation under cyclic
variations in the temperature. As a result, the transformation
follows the same path in cooling and in heating, so that a two-way
(reversible) shape memory effect emerges. There are several ways
in which a reversible shape memory effect can be achieved: through
strains that lie outside the limits of reversible strain in the
austenitic or martensitic state [9, 10]; through multiple
repetition of the cooling--deformation--heating cycles that lead
to the shape memory effect [11]; through multiple repetition of
the loading cycles at a temperature above $A_f$ (the final
temperature of the reversible transformation) that lead to
hyperelasticity (pseudoelasticity) [11]; through repetition of the
cycles "heating above $A_f$ --- cooling under load to $M_f$ (the
final temperature of the martensitic transformation) --- removing
the load --- heating above $A_f$" [12]; and through thermal
treatment under load. All these treatments lead to the emergence
of spontaneous strain upon cooling, and the value of this strain
is smaller than in the ordinary shape memory effect. Moreover, the
stresses generated in the direct transformation are much smaller
than those characteristic of the reverse transformation.
Reversibility disappears even under a small load, although without
a load it can be reproduced several hundred times. The explanation
of the easy disappearance of the shape memory effect in a direct
transformation under load is that for a direct transformation
there are several crystallographic variants, and the variant
chosen is the one that reduces the external stresses. In view of
these crystallographic factors, it is practically impossible to
increase the stress that appears in the direct transformation.

All these effects, which are realized in the stress \textit{vs}
temperature plane, have been thoroughly studied for various
classes of materials and form the basis of our discussion. The
subject of the present review is a description of martensitic
transformations and the accompanying shape memory effects in
ferromagnets. The presence of a magnetic subsystem in substances
that undergo such transformations enriches the picture and makes
it possible to study the phenomena that occur in them using three
coordinates: load, magnetic field strength, and temperature. The
effect of a magnetic field does not amount solely to a
modification of these effects -- it opens up new possibilities in
controlling the shape and size of shape memory ferromagnets. A
magnetic field can be used to shift the temperatures of structural
phase transitions and to affect the topology of the martensitic
phase. Here the parameters of the magnetic subsystem of a
ferromagnet play the leading role. The difference in the
magnetization of austenite and martensite determines the size of
the shift in the temperature of a phase transition in a magnetic
field. The magnetoelastic coupling and magnetocrystalline
anisotropy constants determine the possibility of transforming the
martensitic variants by applying a magnetic field.

The same phenomena that are observed in nonmagnetic materials in
the load \textit{vs} temperature plane can, in principle, be
realized in the magnetic field strength \textit{vs} temperature
plane. Magnetically induced strains that are greater than simple
single-ion magnetostriction have been observed in some Heusler
alloys and in a number of intermetallic compounds. The Heusler
alloy Ni$_2$MnGa, in which the martensitic transformation takes
place in the ferromagnetic state [13], has so far attracted the
greatest attention.

\begin{figure}[t]
\includegraphics[width=5cm]{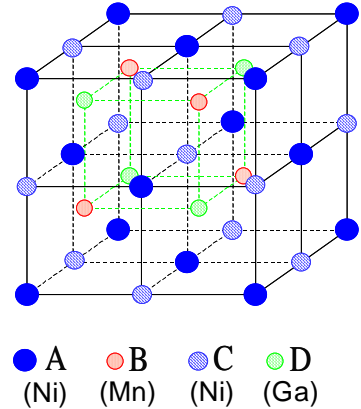}
\caption{L$2_1$ structure of the austenitic phase of Ni$_2$MnGa.}
\end{figure}

\bigskip

\section{Heusler alloy Ni$_2$MnGa: elastic subsystem}

\subsection{Crystal structure}

Heusler alloys are ternary intermetallic compounds with the
general formula X$_2$YZ. The Ni$_2$MnGa alloy, which belongs to
this family, has the L$2_1$ structure at room temperature, which,
as shown in Fig.~1, can be represented by a bcc lattice in which
the Ni atoms occupy the position at the center of the cube, while
the Mn and Ga atoms alternatively occupy the positions at the
apexes [14]. The formation of such a structure from the melt (the
melting point of Ni$_2$MnGa is roughly 1380~K) is, in principle,
possible either from the fully disordered phase A2 $(\mathrm{A}2
\to \mathrm{L}2_1)$ or through the partially ordered intermediate
phase $\mathrm{B}2^{\prime}$ $(\mathrm{A}2 \to
\mathrm{B}2^{\prime} \to \mathrm{L}2_1)$, in which the Ni atoms
already form the frame of the lattice, while the Mn and Ga atoms
still occupy arbitrary positions. But with Ni$_2$MnGa the
situation is different [15]. As the temperature decreases, this
compound passes from the melt directly into the partially ordered
phase B$2^{\prime}$, and this phase then experiences a
second-order phase transition of the disorder--order type [16].
The $\mathrm{B}2^{\prime} - \mathrm{L}2_1$ transition temperature
for Ni$_2$MnGa is about 1070~K. Down to $T_m \sim 200$~K
Ni$_2$MnGa remains in the L$2_1$ phase, and this Heusler alloy
then undergoes a first-order phase transition to a martensitic
tetragonal phase, with $c/a < 1$. At room temperature the cubic
lattice constant $a$ of Ni$_2$MnGa is 5.825~\AA~ and the unit cell
volume $V_{cub} \sim 198$~\AA$^3$ (the number of formula units per
unit cell is $Z = 4$). At low temperatures the parameters of the
tetragonal lattice are $a = b = 5.920$~\AA~ and $c = 5.566$~\AA,
with $c/a = 0.94$, and the unit cell volume is 195~\AA$^3$
[14].$^1$

\footnotetext[1]{Wedel \textit{et al}. [57] suggested an
alternative description of the structure of this compound in the
martensitic phase. They interpreted it as a tetragonal phase of
$I_4/mmm$ symmetry with the crystal lattice parameters $a = 4.18$
\AA~ and $c = 5.56$ \AA, with $c/a = 0.75$, and an unit cell
volume of 97.46 \AA$^3$ (the number of formula units per unit cell
is $Z = 2$). Bearing in mind that in the old [14] and new [57]
descriptions, the values of $a$ differ by a factor of $\sqrt{2}$,
the $c$ parameter is the same, and the volumes per formula unit
coincide, in the following analysis of the experimental data we
will use the crystal orientations accepted in the original works.}

This does not exhaust the sequence of phase transitions in
Ni$_2$MnGa, since these alloys can experience a premartensitic
transition and intermartensitic transformations.

Note that in studies of the crystal structure of the
non-stoichiometric Ni$_2$MnGa alloys by the X-ray (or electron)
diffraction method no reflections corresponding to the L$2_1$
structure have been observed because the atomic scattering factors
of the constituent elements are close to each other. Although the
reflections observed in X-ray diffraction patterns represent only
the short-range order corresponding to the B2 structure,
traditionally the crystal structure of the cubic phase of the
family of Ni$_{2+x+y}$Mn$_{1-x}$Ga$_{1-y}$ alloys is considered as
L$2_1$ structure. It must also be noted that for samples of
non-stoichiometric composition a martensitic phase with
orthorhombic and monoclinic distortions has been reported (see
Section~3.3).

Before we discuss the physical properties of Ni$_2$MnGa any
further, a remark concerning the chemical composition of this
Heusler alloy is in order. Almost all parameters of Ni$_2$MnGa
have proven to be very sensitive to the chemical composition of
the samples. The sample's composition strongly affects the
temperatures of phase transformations and the formation of
superstructures in the austenitic and martensitic states. In many
original publications, the exact composition of the sample is not
given and is assumed to be stoichiometric. This has led to a
situation in which, for instance, the temperatures (known from the
literature) of the martensitic transformation vary from less than
4.2 K to 626 K [17]. Bearing all this in mind, we will give, where
possible, not only the nominal composition but also the
temperatures of the investigated phase transition. The fact that
these parameters do not agree with each other may mean, for one
thing, that the chemical composition of the investigated samples
does not agree with the component ratio in the working mixture.
The dependence of the most important parameters of the Ni$_2$MnGa
alloy on composition is discussed in Section~5.

\subsection{Premartensitic phenomena}

Various pretransitional phenomena occur prior to the structural
transformations of the martensitic type observed in a broad class
of materials. Among these are the formation of soft modes in the
lattice, anomalous broadening of the reflections in the X-ray
spectrum, the emergence of a tweed structure, etc. Pretransitional
phenomena are observed in superconductors with an A15 structure
and in ferroelectrics with a perovskite structure [5] and in a
broad class of shape memory alloys [6, 18]. What sets the Heusler
alloy Ni$_2$MnGa apart from all other compounds that experience
martensitic transformations is that, in addition to
pretransitional phenomena [19, 20], there can be a premartensitic
phase transition.

The physical nature of the premartensitic transition has been the
topic of theoretical [21, 22] and experimental work, including
neutron diffraction [23--25], transport [26, 27], magnetic [28],
mechanical [29--31], and ultrasonic measurements [32--35], and
electron microscopy [36, 37].

The focus in the inelastic neutron scattering study [20] of a
Ni$_2$MnGa single crystal with a martensitic transition $T_m \sim
220$~K was on the transverse acoustic phonon TA$_2$ mode in the
$[\zeta \zeta 0]$ direction. At $\zeta = 0$ this mode corresponds
to the elastic constant $C^{\prime} = (C_{11} - C_{12}/2$, which,
in turn, is determined by the speed of transverse sound
propagating in the [110] direction and polarized along the $[1\bar
10]$ axis. In contrast to the longitudinal LA mode and the
transverse $[\zeta 00]$ TA mode, the dispersion curve of the
$[\zeta \zeta 0]$ TA$_2$ mode exhibits substantial softening at
$\zeta _0 = 0.33$. The development of this feature with decreasing
temperature is shown in Fig.~2a.

The temperature dependence of the squared frequency of the soft
phonon mode is shown in Fig.~2b. This dependence is non-monotonic,
with the frequency of the TA$_2$ mode reaching its minimum (but
not vanishing) at the premartensitic transition temperature $T_P
\sim 260$~K and increasing, as temperature is decreased further.
Since $T_P$ is much higher than the martensitic transformation
temperature $T_m$, this means that in the interval from $T_P$ to
$T_m$ the Ni$_2$MnGa crystal is in an intermediate phase between
austenite and martensite.

The unconventional temperature dependence of the soft phonon mode
indicates that tendencies toward the formation of a micromodulated
superstructure in the Ni$_2$MnGa lattice develop long before the
transition to the martensitic state. Electron microscopy studies
suggest that the overall cubic symmetry of the crystal lattice is
conserved in the interval from $T_m$ to $T_P$ but that in this
lattice a modulation with a period of six atomic separations
develops in the [220] direction, $d_{220} \approx 2$ \AA~[30, 37].

Further development of these neutron diffraction studies [20, 23]
was the work done by Zheludev and Shapiro [24], who studied the
effect of uniaxial stress on the transverse phonon $[\zeta \zeta
0]$ TA$_2$ mode in Ni$_2$MnGa. The uniaxial stress was applied
along the crystallographic [001] axis. In the unstressed sample
the minimum in the dispersion curve was also achieved at $[\zeta
_0 \zeta _00]$, $\zeta _0 = 0.33$. As the stress increased, the
minimum in the dispersion curve shifted toward higher values, and
at $\sigma = 95$~MPa the minimum was at $\zeta _0 = 0.36$. The
temperature dependence of the energy of the anomalous phonons was
found to be nonmonotonic. The greatest softening of the mode was
achieved at $T \sim 300$~K, and with a further decrease in
temperature the energy of the anomalous phonons grew. This
indirect indication of the rise in the premartensitic transition
temperature was corroborated by data on elastic neutron
scattering.

\begin{figure}[h]
\includegraphics[width=7cm]{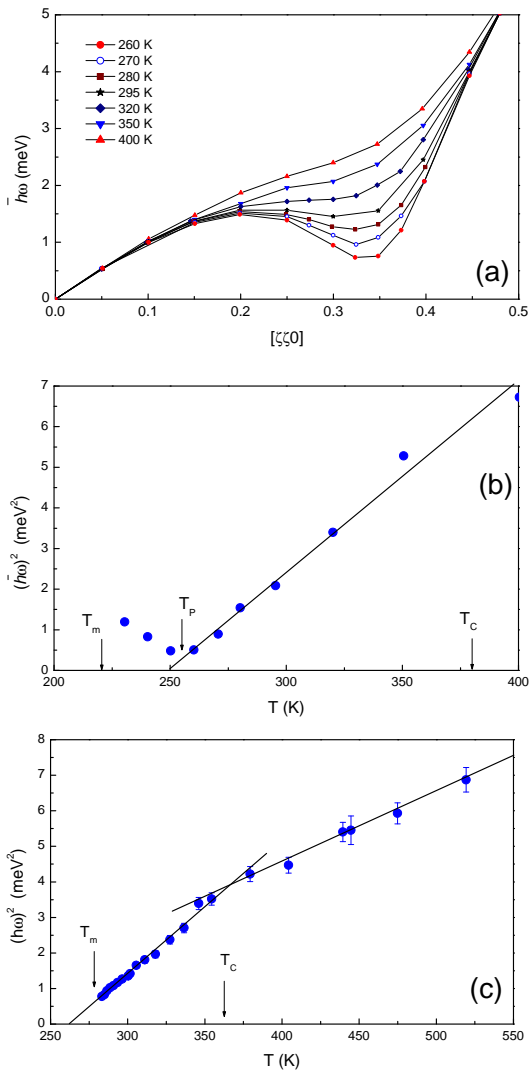}
\caption{(a) Dispersion curves of the phonon $[\zeta \zeta 0]$
TA$_2$ mode in Ni$_2$MnGa [20], (b) temperature dependence of the
squared frequency of the phonon 0.33[110] TA$_2$ mode in
Ni$_2$MnGa [20], and (c) temperature dependence of the squared
frequency of the phonon 0.33[110] TA$_2$ mode in
Ni$_{51.5}$Mn$_{23.6}$Ga$_{24.9}$ [38].}
\end{figure}

On the microscopic level the softening of the $[\zeta \zeta 0]$
TA$_2$ mode at a certain value $\zeta _0$ means that $[\zeta
_0\zeta _00]$ is a special wave vector of the Fermi surface, and
the fact that these vectors coincide in the phonon and electron
subsystems (this is known as nesting) of the metal leads to an
enhancement of the electron--phonon interaction. When pressure is
applied to the sample, the parameters of the unit cell change,
which leads to deformation of the Fermi surface. Nesting in this
case is achieved at another value of the wave vector, and the
anomaly in the phonon dispersion curve shifts to another value of
$\zeta$ [24].

Neutron diffraction studies of Ni$_2$MnGa single crystals with
substantial deviations from stoichiometry have identified several
interesting features of these alloys. The temperature dependence
of the squared frequency of the phonon mode in a sample of
Ni$_{51.5}$Mn$_{23.6}$Ga$_{24.9}$ composition ($T_m = 284$~K and
$T_C = 364$~K) at $\zeta = 0.33$ [38] is depicted in Fig.~2c.
Clearly, at $T_m$ the soft acoustic phonons have a finite energy,
and the squared phonon energy increases linearly with temperature
up to the Curie point $T_C$. For $T > T_C$ the dependence is also
linear, but its slope is much smaller: at $T < T_C$ the slope is
0.039 meV$^2$K$^{-1}$, while at $T > T_C$ it is 0.019
meV$^2$K$^{-1}$. The temperatures to which these lines are
extrapolated from the ferromagnetic and paramagnetic phases are
264~K and 175~K, respectively. All this points to a substantial
effect of magnetic ordering on the elastic subsystem of
Ni$_2$MnGa. Softening of the phonon mode with the same wave vector
was also observed in a sample of Ni$_{52.9}$Mn$_{26.6}$Ga$_{20.5}$
composition with close temperatures of the structural ($T_m =
351$~K) and magnetic ($T_C = 360$~K) phase transitions [39]. In
contrast to the samples whose compositions were close to
stoichiometric, which were studied by Zheludev \textit{et al.}
[20, 23], where the energy of the soft mode was found to increase
as $T_m$ was approached, in the samples with compositions
Ni$_{51.5}$Mn$_{23.6}$Ga$_{24.9}$ and
Ni$_{52.9}$Mn$_{26.6}$Ga$_{20.5}$ no such effect was observed.
This indicates that in the samples in question there was no
premartensitic transition.

Ultrasonic measurements have demonstrated that the temperature
dependencies of the elastic moduli are qualitatively similar to
the temperature dependence of the phonon modes discovered in
neutron diffraction studies. For instance, in Ref. [40] it was
reported that in non-stoichiometric Ni$_2$MnGa the elastic
constants exhibit an anomalous temperature dependence at $T_m =
275$~K approached from higher temperatures; precisely, $C_{12}$
increases and $C_{11}$ and $C_{44}$ decrease, pointing to a
softening of the shear modulus $C^{\prime} = (C_{11} - C_{12}/2$,
which characterizes the stability of the lattice under shear
deformation. The anomalously small value of $C^{\prime}$ is a
distinctive feature of many alloys undergoing martensitic
transitions [41], and, in addition to the tweed structure,
broadening of the reflections in the X-ray spectrum and the soft
phonon mode, is one of the pretransitional phenomena. No
premartensitic transition was observed in the sample studied in
Ref. [40], apparently because of the relatively high martensitic
transition temperature. Ultrasonic measurements involving
single-crystal samples with smaller deviations from stoichiometry
have shown the presence of additional anomalies in the temperature
dependencies of the elastic constants, anomalies that owe their
emergence to a premartensitic transition.

Anomalies in the temperature dependencies of the elastic constants
have been observed at the premartensitic transition temperature
$T_P = 230$~K for a sample whose composition is close to
stoichiometric ($T_m = 175$~K and $T_C = 381$~K) [33]. Figure~3a
shows that the substantial softening of the shear moduli $C_{44}$
and $C^{\prime}$ occurs as we move closer to $T_P = 230$~K.
Further cooling only increases these moduli. Such behavior of the
elastic constants, especially $C^{\prime}$, is similar to the
temperature dependence of the phonon $[\zeta \zeta 0]$ TA$_2$ mode
at $\zeta = 0.33$ [20]. Measurements of the speed and damping of
ultrasound involving a Ni$_2$MnGa single crystal with $T_m =
220$~K and $T_P = 265$~K [32, 34] have shown that the elastic
properties of this sample are qualitatively similar, within a
broad temperature interval, to those of a single crystal with
lower temperatures of the premartensitic and martensitic
transitions studied by Ma\~nosa \textit{et al.} [33].

\begin{figure}[h]
\includegraphics[width=7cm]{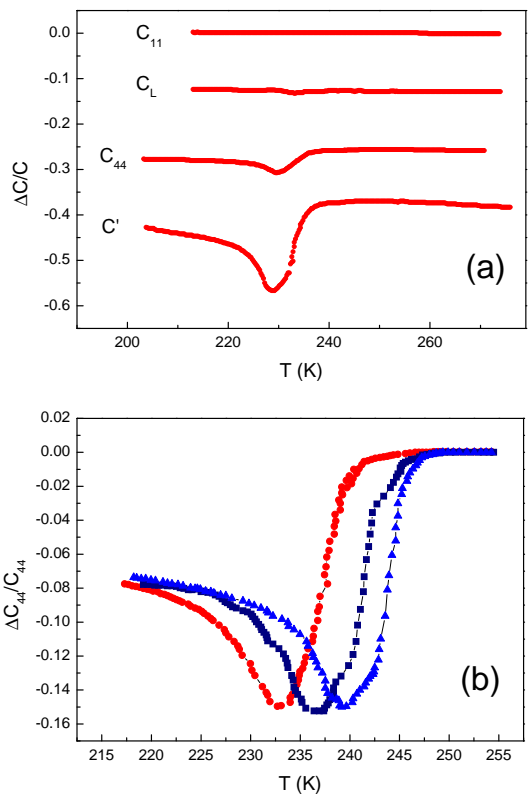}
\caption{(a) Temperature dependencies of the elastic constants in
single crystal Ni$_2$MnGa with $T_m = 175$~K and $T_P = 230$~K
[33], and (b) temperature dependence of the elastic constant
$C_{44}$ under the following uniaxial loads (in units of MPa):
0~($\bullet$), 1~($\blacksquare$), and 4.5~($\blacktriangle$)
[35].}
\end{figure}

The investigations of Gonz\`alez-Comas \textit{et al.} [35] of the
effect of uniaxial compression on the temperature of the
premartensitic transition in Ni$_{49.5}$Mn$_{25.4}$Ga$_{25.1}$
with $T_m = 175$~K and $T_P = 230$~K have revealed that the
minimum in the temperature dependence of the elastic modulus
$C_{44}$ shifts toward higher temperatures when a load is applied
along the crystallographic direction $[1\bar 10]$ (Fig.~3b). Note
that this experiment revealed the presence of pronounced
hysteresis phenomena ($\sim 8$~K at 9~MPa) at the premartensitic
transition temperature. In the absence of uniaxial compression no
temperature hysteresis was observed near $T_P$. The fact that
hysteresis phenomena appear in the premartensitic transition [35]
agrees with the data on the effect of uniaxial compression on the
soft phonon $[\zeta \zeta 0]$ TA$_2$ mode ($\zeta = 0.33$) [24].
An interpretation of this effect was done by Ma\~nosa \textit{et
al.} [42], who found that external pressure applied to the sample
results in the singularities characteristic of first-order phase
transitions becoming more pronounced.

Measurements of the magnetic-field dependencies of the elastic
constants $C_L$, $C_{44}$, and $C^{\prime}$ for single-crystal
Ni$_{49.5}$Mn$_{25.4}$Ga$_{25.1}$ in magnetic fields up to 10~kOe
applied along the crystallographic directions [001] and [110]
revealed [35] that the elastic constants increase in the magnetic
field and reach saturation in fields $\sim 1$~kOe for \textbf{H}
$\Vert$ [001]. For \textbf{H} $\Vert$ [110], saturation was
reached at $\sim 3$~kOe. The results of magnetization measurements
done with the same sample [43] suggest that the variation of the
elastic constants in a magnetic field is proportional to the
square of magnetization, $M^2$.

\subsection{Superstructural motives}

Almost all X-ray and neutron diffraction studies involving
Ni$_{2}$MnGa reveal the presence of superstructural reflections,
in addition to the main reflections of the low-temperature
martensitic phase. For instance, in the first neutron diffraction
studies of a sample of stoichiometric composition [14] it was
found that there are additional reflections of the tetragonal
phase and it was assumed that this phase is modulated along,
apparently, the [100] direction. Further studies of the crystal
structure of the low-temperature phase in non-stoichiometric
compounds revealed a complex pattern of formation of different
martensitic phases and the presence of intermartensitic phase
transitions in the system Ni$_{2+x+y}$Mn$_{1-x}$Ga$_{1-y}$
[44-57]. In the early stages of such studies the superstructural
motifs were described as static displacement waves (modulations)
[44], although lately an alternative approach is being developed,
in which the superstructural reflections of martensite are
interpreted as long-period rearrangements of closely packed planes
of the $\{100\}$ type [25, 57]. A comparative analysis of these
two approaches in describing the crystal structure of martensite
in Ni-Mn-Ga is given in Ref. [56], where it is shown that they
often lead to the same results.

As of now, the existence of an unmodulated martensitic phase and
martensite with five- and seven-layered modulations along the
crystallographic direction [110] has been established. There have
also been reports about observations of longer-period modulations
and about intermartensitic transformations in
Ni$_{2+x+y}$Mn$_{1-x}$Ga$_{1-y}$ induced by temperature or
uniaxial strain.

Five-layered modulation of the low-temperature tetragonal phase
was observed by the method of diffraction of electrons and X-rays
on single crystals of Ni$_{51.5}$Mn$_{23.6}$Ga$_{24.9}$ ($M_s =
293$~K) [44, 48, 53], Ni$_{49.2}$Mn$_{26.6}$Ga$_{24.2}$ ($M_s \sim
180$~K) [51, 52], Ni$_{52.6}$Mn$_{23.5}$Ga$_{23.9}$ ($M_s =
283$~K) [51], Ni$_{52}$Mn$_{23}$Ga$_{25}$ ($M_s = 227$~K) [25],
and Ni$_{48.5}$Mn$_{30.3}$Ga$_{21.2}$ ($M_s = 307$~K) [58]. In the
process of formation of superstructures in the martensitic phase
of these Heusler alloys, the X-ray patterns showed, besides the
main diffraction reflections, a number of additional reflections.
Modulation occurs in such a way that each fifth (110) plane does
not undergo displacements, while the other four are displaced from
their regular positions in the body-centered tetragonal lattice
along the [110] direction.

Seven-layered modulation of the martensitic phase was observed in
single crystals of Ni$_{52}$Mn$_{25}$Ga$_{23}$ ($M_s = 333$~K)
[44] and Ni$_{48.8}$Mn$_{29.7}$Ga$_{21.5}$ ($M_s = 337$~K) [59].
X-ray studies of Ni$_{48.8}$Mn$_{29.7}$Ga$_{21.5}$ have shown
that, as in the case of five-layered martensite, besides the main
reflections there are additional diffraction reflections that lie
along the [110] direction. The crystal structure of the
seven-layered martensite in Ni$_{48.8}$Mn$_{29.7}$Ga$_{21.5}$ was
found to be orthorhombic with the lattice parameters $a =
6.19$~\AA, $b = 5.80$~\AA, and $c = 5.53$~\AA~ [59]. In contrast
to this, the crystal structure of seven-layered martensite in
Ni$_{52}$Mn$_{25}$Ga$_{23}$ was interpreted as monoclinic with the
lattice parameters $a = 6.14$~\AA, $b = 5.78$~\AA, and $c =
5.51$~\AA~ and with $\gamma = 90.5^{\circ}$ [44].

Unmodulated martensite was observed in single-crystal samples and
thin films of alloys of the following compositions:
Ni$_{53.1}$Mn$_{26.6}$Ga$_{20.3}$ ($M_s = 380$~K) [54, 60] and
Ni$_{48.5}$Mn$_{30.3}$Ga$_{21.2}$ ($M_s = 307$~K) [58]. The
crystal structure of the unmodulated martensitic phase proved to
be tetragonal, but the ratio $c/a \sim 1.18$ for it is much higher
than for the tetragonal martensitic phases considered above.

The experimental data on modulations of the crystal lattice in the
martensitic phase suggest that the type of martensite (modulation
period) depends on the composition of the alloys. It is convenient
to classify these types according to the martensitic transition
temperature $M_s$ [56]. Five-layered martensite has a crystal
lattice of tetragonal symmetry and is formed in the process of
cooling in alloys that have $M_s < 270$~K. Seven-layered
martensite is formed in alloys with a higher $M_s$ and has a
crystal structure different from the tetragonal one.

\section{Heusler alloy Ni$_2$MnGa: magnetic subsystem}

\subsection{Magnetization}

The results of band structure calculations for the Heusler alloys
$X_2$Mn$Z$ [61, 62] suggest that the Mn spin-up $3d$-states are
almost completely occupied and that the Mn $3d$-wave functions
overlap with the $3d$-wave functions of the $X$ atoms. The same
calculations show that Mn spin-down $3d$-states are almost
entirely unoccupied. Recently these calculations have been
corroborated by the data from photoemission spectroscopy [63-65].
Webster \textit{et al.} [14] showed that the magnetic moment in
the Heusler alloy Ni$_2$MnGa is localized primarily on the
manganese atoms, $\mu _{\mathrm{Mn}} \approx 4.17\mu _B$, while
the magnetic moment on nickel atoms $\mu _{\mathrm{Ni}} \le 0.3\mu
_B$.

\begin{figure}[h]
\includegraphics[width=7cm]{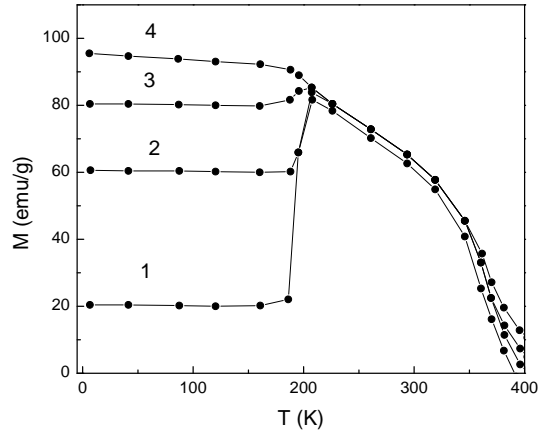}
\caption{Temperature dependencies of magnetization of Ni$_2$MnGa.
1 -- 1~kOe, 2 -- 4~kOe, 3 -- 8~kOe, and 4 -- 16~kOe [14].}
\end{figure}

The temperature dependencies of the magnetization of Ni$_2$MnGa
measured in fields ranging from 1~kOe to 16~kOe [14] are shown in
Fig.~4. Clearly, this alloy transforms to the ferromagnetic state
at the Curie temperature $T_C = 376$~K. In the martensitic
transformation region at $T_m = 202$~K in weak fields there is a
sharp decline in the magnetization, which corresponds to an
increase in magnetocrystalline anisotropy, while measurements in
strong fields have shown that the saturation magnetization in the
martensitic phase is somewhat greater than the saturation
magnetization in the austenitic phase.

The reports of systematic studies of the effect of deviation from
the stoichiometric composition on the magnetic properties of
Ni$_{2+x}$Mn$_{1-x}$Ga can be found in Refs [66-68]. Substitution
of Ni atoms for Mn atoms is accompanied by a decrease in
magnetization, an increase in $T_m$, and a decrease in $T_C$. The
decrease in the Curie temperature and magnetization can be
explained by the decrease in the number of atoms that carry the
magnetic moment, while the increase in the martensitic
transformation temperature is, probably, caused by the increase in
the number of conduction electrons. Nickel atoms have three more
electrons in the $d$-shell than manganese atoms. Substitution of
Ni atoms for Mn atoms leads to an increase in the volume of the
Fermi surface and thus to an increase in the temperature at which
the structural phase transition takes place. Note that the
martensitic transformation disappears even when a small number
($\sim 10$\%) of Mn atoms are replaced by V atoms [69].

In addition to the anomalies caused by the martensitic
transformation, specific features in the temperature and field
dependencies of the magnetic properties of Ni$_2$MnGa caused by
intermartensitic and premartensitic transitions have also been
observed. The anomalies associated with intermartensitic
transitions manifest themselves most vividly in the temperature
dependencies of the low-field magnetic susceptibility [49, 59,
70], which is an indication that the magnetocrystalline anisotropy
changes upon the transition of one martensitic phase into another.
For instance, a sequence of intermartensitic transitions has been
observed in the reheating of a Ni$_{52.6}$Mn$_{23.5}$Ga$_{23.9}$
single crystal that was first preloaded along the [110]
crystallographic direction and then unloaded at liquid nitrogen
temperature [49]. Figure~5 shows that, in addition to the
anomalies in the low-field magnetic susceptibility $\chi$ at the
Curie temperature $T_C = 370$~K and the martensitic transition
temperature $T_m = 290$~K, there are two more anomalies, at 190~K
and at 230~K. These new anomalies were interpreted as being caused
by intermartensitic transitions from the unmodulated martensitic
phase to seven-layered martensite at 190~K and then from
seven-layered martensite to five-layered martensite at 230~K. The
results of the measurements of magnetization in a 15~kOe field
that Kokorin \textit{et al.} [71] conducted on a single
crystalline sample of the same composition and in the same
experimental conditions (preloading and unloading at 77~K)
suggested that the martensitic phases have different
magnetizations. However, since the change in magnetization was
detected only for one intermartensitic transition, the researchers
assumed that the magnetizations of the other two martensitic
phases have close values.

\begin{figure}[h]
\includegraphics[width=7cm]{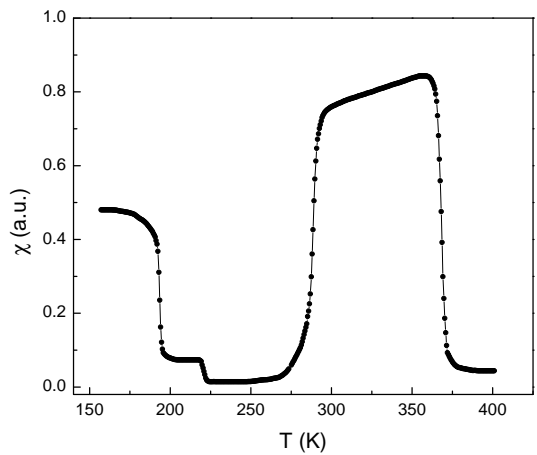}
\caption{Temperature dependence of low-field magnetic
susceptibility of a Ni$_{52.6}$Mn$_{23.5}$Ga$_{23.9}$ single
crystal that was preloaded and then unloaded at liquid nitrogen
temperature [49].}
\end{figure}

Certain anomalies in the vicinity of the premartensitic phase
transition have also been observed at temperature dependencies of
the low-field magnetic susceptibility and magnetization of
Ni$_2$MnGa [27, 28, 33, 72]. Since the formation of the
premartensitic phase is accompanied by the formation of a static
displacement wave [37], we can assume that the magnetic anisotropy
in the premartensitic phase is higher than in the cubic phase,
which is reflected by a drop in magnetization at $T_P$, as shown
in Fig.~6.

\begin{figure}[t]
\includegraphics[width=6cm]{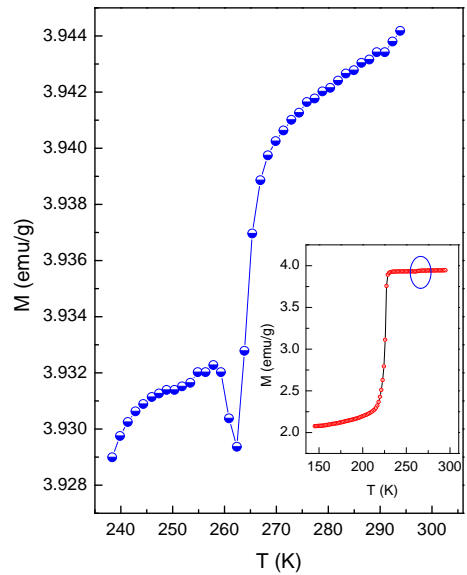}
\caption{Temperature dependence of the magnetization $M$ of
polycrystalline Ni$_{2.02}$Mn$_{0.98}$Ga in a field of 100~Oe in
the vicinity of premartensitic transition. The inset depicts the
behavior of $M(T)$ within a broad temperature range.}
\end{figure}

\begin{figure}[h]
\includegraphics[width=6cm]{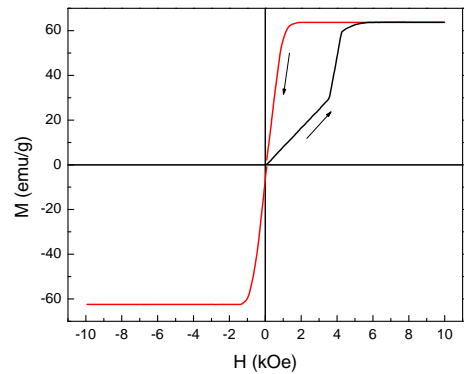}
\caption{Field dependencies of magnetization of the martensitic
phase of the Ni$_{48}$Mn$_{31}$Ga$_{21}$ single crystal at $T =
290$~K [74].}
\end{figure}

The transformation of the martensitic domains with a magnetization
vector oriented unfavorably with respect to the external magnetic
field may lead to anomalies in the field dependencies of
magnetization [73-75]. Figure~7 depicts the field dependencies of
magnetization for a single crystal Ni$_{48}$Mn$_{31}$Ga$_{21}$
sample with the following temperatures of the ferromagnetic and
martensitic transitions: $T_C = 371$~K and $M_s = 300$~K [74].
Measurements done at 290~K have shown that the slow increase in
magnetization is replaced by a rapid increase at a critical value
of the field, $H \approx 3.5$~kOe. The magnetization reaches
saturation in a field of the order 5~kOe. In the subsequent
reduction of field strength, magnetization remains practically the
same down to 1.3~kOe, which results in large hysteresis.
Hysteresis is observed only during the first magnetization cycle,
while during the second cycle the magnetization reaches its
saturation value in a weak magnetic field. The anomalous behavior
of the magnetization curves can be explained by the fact that at a
certain critical field strength the martensitic domains are
redistributed even before the magnetization vector has time to
fully rotate into a position parallel to the magnetic field. The
redistribution of the martensitic domains proceeds in such a way
that the $c$ axis, which is the easy magnetization axis of the
tetragonal martensitic phase, aligns itself with the external
magnetic field.

Although Ni$_{2+x+y}$Mn$_{1-x}$Ga$_{1-y}$ alloys basically exhibit
magnetic properties typical of ferromagnets, some alloys of this
system demonstrate unusual magnetic characteristics, such as, say,
hysteresis of the ferromagnetic transition and "metamagnetic"
anomalies in alloys with the same temperatures of structural and
magnetic phase transitions [76, 77]. Such a situation can be
achieved if some of the manganese atoms are replaced by nickel
atoms or some of the gallium atoms are replaced by nickel atoms.
In the first case the coupled magnetic and structural transition
is realized in Ni$_{2+x}$Mn$_{1-x}$Ga $(x = 0.18 -- 0.20)$ [67,
78], while in the second case it is realized in
Ni$_{53}$Mn$_{25}$Ga$_{22}$ [79]. Hysteresis of the ferromagnetic
transition and the metamagnetic behavior of magnetization near the
magnetostructural transition are an indication that for
Ni$_{2.18}$Mn$_{0.82}$Ga and Ni$_{2.19}$Mn$_{0.81}$Ga the magnetic
transition has the characteristics of a first-order phase
transition [76, 77]. The reason for this is that these alloys
undergo a first-order phase transition from the tetragonal
ferromagnetic phase into the cubic paramagnetic phase.

\subsection{Magnetocrystalline anisotropy}

Magnetocrystalline anisotropy can be considered as the determining
parameter for the achievement of giant strains induced by a
magnetic field in shape memory ferromagnets. Although it seems, at
first glance, that the formation and growth of the structural
domains favorably oriented with respect to the magnetic field are
in no way related to this parameter, it is precisely this
parameter that determines the path along which the system proceeds
and reaches the state with the lowest possible energy.

Measurements of magnetization in a
Ni$_{51.3}$Mn$_{24.0}$Ga$_{24.7}$ single crystal have shown that
the easy magnetization axis in the cubic phase is oriented along
the crystallographic [100] axis and that the magnetocrystalline
anisotropy constant $K_1$ in this phase is relatively moderate. As
a result of the transition to the martensitic phase the
magnetocrystalline anisotropy changes significantly. The field
dependencies of the magnetization of
Ni$_{51.3}$Mn$_{24.0}$Ga$_{24.7}$ in the martensitic phase are
shown in Fig.~8a. In these curves one can isolate intervals within
which the displacement of the domain walls and rotation of the
magnetization vectors in the domain occur.

\begin{figure}[t]
\includegraphics[width=6cm]{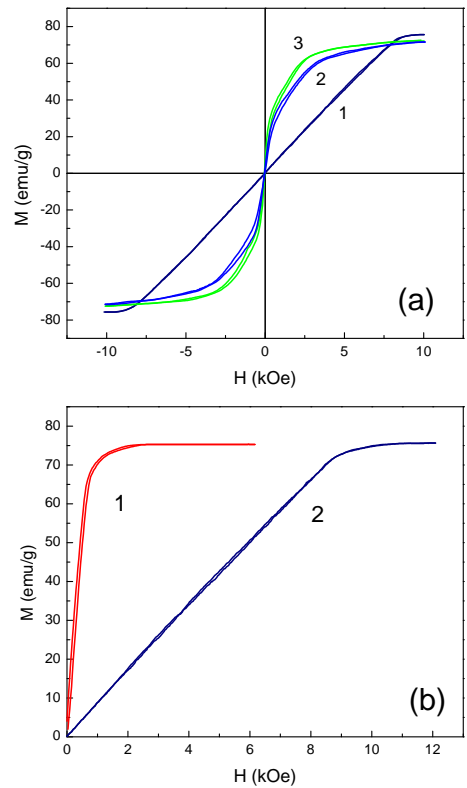}
\caption{(a) Field dependencies of magnetization of the
Ni$_{51.3}$Mn$_{24.0}$Ga$_{24.7}$ single crystal for a
multivariant martensitic state; 1 -- \textbf{H} $\parallel$ [100],
2 -- \textbf{H} $\parallel$ [110], and 3 -- \textbf{H} $\parallel$
[111]. The orientation of the single crystal was done in the
austenitic phase. (b) Field dependencies of magnetization of the
Ni$_{51.3}$Mn$_{24.0}$Ga$_{24.7}$ single crystal for a
single-variant martensitic state: 1 -- easy magnetization axis,
and 2 -- hard magnetization axis [80].}
\end{figure}

The redistribution of the martensitic variants that is induced by
a magnetic field is superimposed on the intrinsic magnetization
processes. To exclude this factor, Tickle and James [80] employed
the following procedure in their measurements of
magnetocrystalline anisotropy.

A thin single crystal plate measuring $5.2 \times 5 \times 0.64$
mm$^3$ whose faces coincided with the crystallographic planes of
the $\{100\}$ type was fabricated for the measurements. The
strategy of the experiment consisted in compressing the sample to
a single-variant state in a spring minipress [80]. Measurements of
the $M$ \textit{vs} $H$ dependence were done in a magnetic field
that was parallel or perpendicular to the direction along which
the sample was compressed. The sample was first cooled below the
martensitic transformation temperature in a 6~kOe magnetic field
directed along the compression axis in such a way that both the
magnetic field and the applied stress facilitated the formation of
a single variant of martensite. The sample size in the direction
of compression was selected in such a way that in the martensitic
phase the sample would be of a square shape. In this case the
demagnetization factors for two orientations of the magnetic field
proved to be the same, and the uniaxial anisotropy constant $K_u$
could be evaluated by the area between the curves. The field
dependencies of the magnetization of the single-variant martensite
(Fig.~8b) show that the $c$ axis is indeed the easy magnetization
axis and that the curves demonstrate uniaxial anisotropy.

The results of measurements under a 1.9~MPa load were used to
calculate the areas between the $M$ axis and the curves of
magnetization along the easy and hard axes. The uniaxial
anisotropy constant in the martensitic phase calculated in this
manner $(K_u = 2.45 \times 10^6$ erg/cm$^{-3}$ at $T = 256$~K)
proved to be greater than the constant $K_1$ in the austenitic
phase by a factor of 100.

Studies of the magnetocrystalline anisotropy of single crystals
Ni$_{2+x+y}$Mn$_{1-x}$Ga$_{1-y}$ of different compositions (the
samples were converted to a single-variant state) have shown that
the values of $K_u$ at room temperature vary from $1.7 \times
10^6$ erg/cm$^{-3}$ for Ni$_{48}$Mn$_{31}$Ga$_{21}$ [74] to $2.48
\times 10^6$ erg/cm$^{-3}$ for Ni$_{49.7}$Mn$_{28.7}$Mn$_{21.6}$
[81].

The temperature dependencies of the uniaxial magnetocrystalline
anisotropy constant for martensite have been measured for
polycrystalline Ni$_2$MnGa [82] and for a
Ni$_{48.8}$Mn$_{28.8}$Ga$_{22.6}$ single crystal [83]. For the
polycrystalline sample it was found that $K_u = 2.5 \times 10^6$
erg/cm$^{-3}$ at $T = 220$~K. As the temperature is lowered, the
value of $K_u$ increases linearly and at 77~K reaches the value of
$3.8\times 10^6$ erg/cm$^{-3}$. The increase in $K_u$ with
decreasing temperature has also been observed for a single-variant
single crystalline sample of Ni$_{48.8}$Mn$_{28.6}$Ga$_{22.6}$
composition. At 283~K the magnetocrystalline anisotropy constant
$K_u = 2 \times 10^6$ erg/cm$^{-3}$, while at 130~K its value
proved to be equal to $2.65 \times 10^6$ erg/cm$^{-3}$. The
theoretical analysis of magnetocrystalline anisotropy done by
Enkovaara \textit{et al.} [84] showed that magnetic anisotropy
changes sign at $c/a = 1$, which corresponds to a change in the
easy magnetization axis from [100] for $c/a < 1$ to [110] for $c/a
> 1$.

Just as the other physical parameters, the magnetocrystalline
anisotropy of the Ni$_{2+x+y}$Mn$_{1-x}$Ga$_{1-y}$ alloys varies
with the chemical composition of these compounds. The lack of
experimental data, however, does not allow us to express this
dependence more specifically.

\subsection{Structure of magnetic domains}

The structure of the magnetic domains in
Ni$_{2+x+y}$Mn$_{1-x}$Ga$_{1-y}$ has been studied in single
crystalline thin films [85-87] and bulk single crystals [88]
faceted along planes of the $\{100\}$ type. The results of these
studies show that the domain structure of the austenitic phase is
formed primarily by $180^{\circ}$ domains with magnetization
vectors \textbf{M} $\parallel$ [100]. The magnetic domain
structure changes dramatically upon martensitic transformation,
when martensitic variants are formed. In the low-temperature phase
several magnetic domains are located within a single martensitic
variant. Since adjacent martensitic variants are separated by a
twin boundary, this leads to the appearance of a relief on the
sample's surface, and the magnetization vectors in adjacent
martensitic variants prove to be directed at a certain angle with
respect to each other [85]. The domain structure within a single
martensitic variant consists of $180^{\circ}$ domains, just as in
the austenitic phase. The width of magnetic domains varies from 5
to $40~\mu \mathrm{m}$ [88]. When a magnetic field is applied to
the sample, the ferromagnetic and martensitic domain structures
change. A weak field changes the topology of the ferromagnetic
domains which come to resemble a "herring-bone" structure with a
common domain wall coinciding with the twin boundary [85]. As the
magnetic field strength is increased further, the processes of
reorientation of the magnetic moment in the ferromagnetic domains
and of displacement of the boundaries between the martensitic
variants begin to compete.

\section{Dependence of properties on composition}

Systematic studies of the ferromagnetic shape memory alloys
Ni$_2$MnGa have shown that magnetic and structural transitions
occur not only in the stoichiometric alloy but also in cases with
significant deviations from stoichiometry [17]. Hence it is only
natural to examine the entire system of Heusler alloys
Ni$_{2+x+y}$Mn$_{1-x}$Ga$_{1-y}$ in which shape memory can be
magnetically controlled. Obviously, finding the composition
dependencies of the main physical properties is important in order
to effectively use these materials in applications. Since the
compositions of studied Ni$_{2+x+y}$Mn$_{1-x}$Ga$_{1-y}$ alloys
are located on the triple phase diagram of this system in an
arbitrary manner, their main properties are considered in the
literature as functions of the average electron concentration per
atom, $e/a$. Such an approach was used by Chernenko [89], who
built an empirical dependence of the martensitic transition
temperature $T_m$ on the electron concentration $e/a$. The
electron concentration in Ni$_{2+x+y}$Mn$_{1-x}$Ga$_{1-y}$ alloys
was calculated using the following electrons configurations of the
outer shells and the following numbers of electrons per atom (in
parentheses): Ni --- 3d$^8$4s$^2$~(10), Mn --- 3d$^5$4s$^2$~(7),
and Ga --- 4s$^2$4p$^1$~(3).

\subsection{Ferromagnetic transition}

In examining the magnetic properties of
Ni$_{2+x+y}$Mn$_{1-x}$Ga$_{1-y}$ alloys, the approach based on
using the total electron concentration $e/a$ has certain
limitations, since the magnetic moment in these alloys is on the
manganese atoms. For instance, $e/a$ can be increased in two ways:
without changing the Mn content (substituting Ni for Ga), or by
changing the Mn content (substituting Ni for Mn). Obviously, the
two ways of changing the electron concentration are not equivalent
in the effect on the magnetic properties of
Ni$_{2+x+y}$Mn$_{1-x}$Ga$_{1-y}$. With the exception of
Ni$_{2+x}$Mn$_{1-x}$Ga $(y = 0)$ alloys, no systematic studies of
the composition dependence of theNi$_{2+x+y}$Mn$_{1-x}$Ga$_{1-y}$
alloys have been done. The dependence of the Curie temperature
$T_C$ on the electron concentration $e/a$ is shown in Fig.~9a for
a large number of Ni$_{2+x+y}$Mn$_{1-x}$Ga$_{1-y}$ alloys [89].
The Curie temperature was found to depend weakly on electron
concentration in the interval $7.3 \le e/a \le 7.6$; at higher
values of $e/a$ the Curie temperature decreases.

\begin{figure}[t]
\includegraphics[width=7cm]{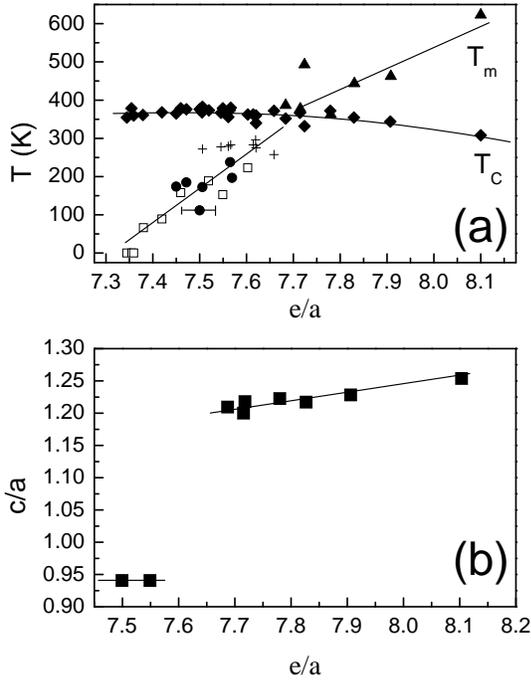}
\caption{(a) Concentration dependencies of the temperatures of
martensitic $(T_m)$ and magnetic $(T_C)$ transitions in
Ni$_{2+x+y}$Mn$_{1-x}$Ga$_{1-y}$ alloys [89], and (b)
concentration dependence of tetragonal distortion of the crystal
structure in Ni$_{2+x+y}$Mn$_{1-x}$Ga$_{1-y}$ alloys [97].}
\end{figure}

The composition dependencies of magnetic properties for
Ni$_{2+x}$Mn$_{1-x}$Ga alloys have been studied in Refs [26, 67,
68, 90, 91]. The phase diagram of these alloys (Fig.~10) clearly
shows that the Curie temperature decreases as Ni is substituted
for Mn in the interval $0 < x < 0.18$; with a further increase in
the Ni content the Curie temperature and the martensitic
transition temperature merge, and an increase in $T_C$ is observed
in the interval $0.18 < x < 0.20$. Wang \textit{et al.} [68]
studied Ni$_{2-x}$Mn$_{1+x/2}$Ga$_{1+x/2}$ alloys $(x = 0-0.1)$
and found that within this concentration range the Curie
temperature lowers from roughly 384~K $(x = 0)$ to roughly 370~K
$(x = 0.1)$. The decrease in the Curie temperature is accompanied
by a decrease in the saturation magnetization $M_0$ and the
magnetic moment on Mn atoms.

\begin{figure}[t]
\includegraphics[width=6cm]{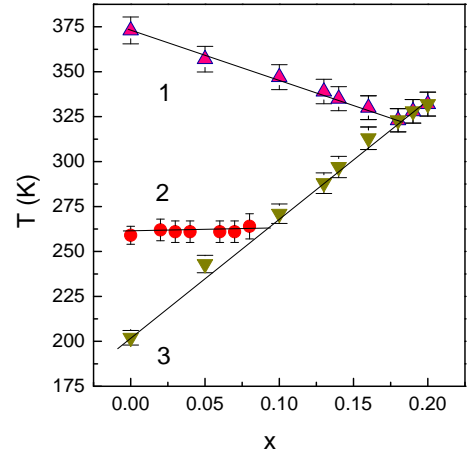}
\caption{Curie temperature $T_C$ (1) and the temperatures of
premartensitic $T_P$ (2) and martensitic $T_m$ (3) transitions in
Ni$_{2+x}$Mn$_{1-x}$Ga alloys as functions of composition.}
\end{figure}

Comparison of the experimental data on the magnetic properties of
Ni$_{2+x}$Mn$_{1-x}$Ga and Ni$_{2-x}$Mn$_{1+x/2}$Ga$_{1+x/2}$
shows that any deviation from stoichiometry results in a decrease
in the Curie temperature and the saturation magnetization. Many
experimental observations, such as the decrease in $T_C$ in
Ni$_{2+x}$Mn$_{1-x}$Ga alloys [67], the increase in $T_C$ in
Ni$_2$Mn$Z$ ($Z$ = Al, Ga, In, Sn, Sb) alloys under pressure [92],
the decrease in $T_C$ upon isoelectronic substitution of In atoms
for Ga atoms, which leads to an increase in the crystal lattice
parameter [93], can be explained by the change in the average
distance between the Mn atoms, the carriers of magnetic moment. At
the same time, the results of Wang \textit{et al.} [68] show that
a simple increase in the number of Mn atoms per formula unit does
not lead to rise in the Curie temperature or to an increase in the
magnetic moment of the alloy.

\subsection{Martensitic transition}

For Heusler alloys the martensitic transition temperature changes
substantially under deviations from stoichiometry [17, 67, 68, 91]
and under doping [69, 89, 93-95]. As a result of their analysis of
the experimental data on Ni$_{2+x+y}$Mn$_{1-x}$Ga$_{1-y}$ alloys,
Chernenko \textit{et al.} [17] concluded that for fixed Mn content
the increase in Ga content lowers the martensitic transition
temperature $T_m$. The same effect can be achieved by substituting
Mn atoms for Ni atoms with the Ga content constant. Substitution
of Mn for Ga with the Ni content constant increases the
martensitic transition temperature. The next step in establishing
the composition dependencies of $T_m$ was done in Ref.~[89], where
the martensitic transition temperature of
Ni$_{2+x+y}$Mn$_{1-x}$Ga$_{1-y}$ alloys was studied as a function
of the electron concentration $e/a$ (Fig.~9a). In alloys with $e/a
\le 7.7$, the martensitic transition temperature increases with
electron concentration with a coefficient of roughly
900~K~$(e/a)^{-1}$, while for alloys with $e/a \ge 7.7$ this
coefficient is of the order 500~K~$(e/a)^{-1}$. In Ref.~[96] the
$e/a$-dependence of $T_m$ is described as $T_m =
[702.5(e/a)-5067]$~K, which makes it possible to define the
empirical dependence of $T_m$ on the molar Mn content
$(X_{\mathrm{Mn}})$ and the molar Ga content $(X_{\mathrm{Ga}})$
as $T_m = (1960 - 21.1X_{\mathrm{Mn}} - 49.2X_{\mathrm{Ga}})$~K.
This approach can also be used in examining the composition
dependence of the martensitic transition temperature $T_m$ in
Ni$_{2+x}$Mn$_{1-x}$Ga alloys $(x = 0-0.2)$ (see Fig.~10), since
substitution of Ni for Mn increases the electron concentration.
The results of Wang \textit{et al.} [68] are an exception to this
empirical formula: they found that under isoelectronic change in
composition in Ni$_{2-x}$Mn$_{1+x/2}$Ga$_{1+x/2}$ $(x = 0-0.1)$
alloys the martensitic transition temperature rises from $\approx
190$~K to $\approx 280$~K in the interval $0 \le x \le 0.06$. With
further increase in $x$ the temperature $T_m$ rapidly decreases
($T_m \approx 200$~K at $x = 0.08$ and $T+m < 100$~K at $x =
0.1$).

\begin{figure}[t]
\includegraphics[width=6cm]{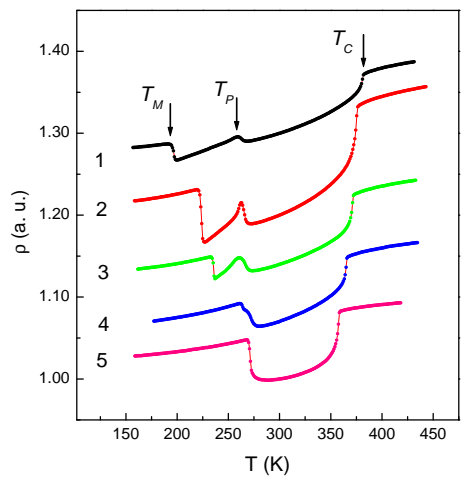}
\caption{Temperature dependence of electrical resistivity in
Ni$_{2+x}$Mn$_{1-x}$Ga alloys for $x = 0$ (curve~1), $x = 0.02$
(2), $x = 0.04$ (3), $x = 0.06$ (4), and $x = 0.09$ (5).}
\end{figure}

\subsection{Premartensitic transition}

As noted in Section~3.2, premartensitic transitions have been
observed in Ni$_{2+x+y}$Mn$_{1-x}$Ga$_{1-y}$ the alloys with
relatively small deviations from the stoichiometry and martensitic
transition temperatures lower than 270~K. For the family of
Ni$_{2+x+y}$Mn$_{1-x}$Ga$_{1-y}$ alloys, the phase diagrams which
represent the temperatures of the magnetic, premartensitic, and
martensitic transitions as functions of $e/a$, show that the
premartensitic transition temperature $T_P$ increases with
electron concentration [31, 35, 72].

Pronounced anomalies at the martensitic transition temperature
$T_P$ have been observed, as Fig.~11 shows, in the temperature
dependencies of the electrical resistivity of
Ni$_{2+x}$Mn$_{1-x}$Ga $(x = 0 - 0.09)$ alloys [27]. The
observation of these anomalies has made it possible to modify the
phase diagram of Ni$_{2+x}$Mn$_{1-x}$Ga alloys so that it allows
for the presence of a modulated premartensitic phase. Figure~10
shows that the premartensitic transition temperature $T_P$ is
weakly dependent on the substitution of Mn atoms for Ni atoms in
Ni$_{2+x}$Mn$_{1-x}$Ga alloys, which leads to a gradual narrowing
of the temperature range within which the premartensitic phase
exists, and the phase completely disappears in the
Ni$_{2.09}$Mn$_{0.91}$Ga $(x = 0.09)$ alloy.

Zheludev \textit{et al.} [23] suggested that premartensitic
transitions emerge because of nesting singularities in the Fermi
surface. Since the premartensitic transition temperature $T_P$ is
weakly dependent on deviations from stoichiometry and such a
transition is observed only in Ni$_{2+x+y}$Mn$_{1-x}$Ga$_{1-y}$
alloys with $T_m < 270$~K [31], variations in conduction electron
concentration have, probably, only a slight effect on the nesting
section of the Fermi surface. In alloys with $T_m > 270$ K,
martensitic transitions, accompanied by a drastic change of the
Fermi surface, occur even before the nesting singularities in the
cubic phase have time to manifest themselves. No premartensitic
transition is observed in this case.

\subsection{Structural distortions}

The data in Refs [56, 97, 98] suggest that the ratio $c/a$, which
gives the tetragonal distortion of the cubic lattice caused by the
transition to the martensitic state, depends on alloy composition.
As Figure~9b shows, $c/a \approx 0.94$ for alloys in which the
electron concentration $e/a \sim 7.4 - 7.6$. An abrupt change in
the tetragonality parameter from $c/a < 1$ to $c/a > 1$ takes
place in alloys with $e/a \ge 7.7$. In contrast to the alloys with
$e/a < 7.6$, for which $T_m < T_C$, in alloys with $e/a \ge 7.7$
the martensitic transition takes place in the paramagnetic state.

The exact position of the "dividing line" for electron
concentration, $e/a = 7.62$, between the tetragonal phases with
$c/a < 1$ and $c/a > 1$ is given in Ref.~[98]. This value is
characteristic of alloys in which there occurs a combined
magnetostructural phase transition, i.e., $T_m \approx T_C$. It
may be assumed that the fact that these two temperatures coincide,
which results in an enhancement of magnetoelastic interaction,
leads to a considerable increase in the degree of tetragonal
distortions of the cubic lattice. In this connection it must be
noted that Ni$_{2.16}$Mn$_{0.84}$Ga ($T_m < T_C$ and $e/a = 7.62$)
and Ni$_{2.19}$Mn$_{0.81}$Ga ($T_m \approx T_C$ and $e/a = 7.64$)
alloys have different mechanical properties and, possibly,
different martensitic transformation kinetics [99].
Polycrystalline samples of the Ni$_{2.19}$Mn$_{0.81}$Ga alloys
have been known (through experiments) to rapidly disintegrate
under thermocycling through the martensitic transition
temperature, while the polycrystalline samples of the
Ni$_{2.16}$Mn$_{0.84}$Ga alloy prepared by the same method endure
multiple thermocycling through $T_m$ without substantial
degradation of mechanical properties. Possibly, the rapid
disintegration of the Ni$_{2.19}$Mn$_{0.81}$Ga samples can be
explained by the fact that these samples undergo larger
distortions (compared to those in Ni$_{2.16}$Mn$_{0.84}$Ga) of the
cubic lattice in the martensitic transformation.

\section{Magnetostrains in Ni$_2$MnGa}

\subsection{Magnetostrain caused by the shift of the martensitic
transition}

The existence of a structural phase transition in the
ferromagnetic matrix opens up new possibilities in magnetically
controlling the temperature of this transition. The extent of this
control is determined by the difference in the magnetizations of
the high- and low-temperature phases. The maximum variation in the
linear dimensions of the sample achieved through shifting the
structural transition temperature is equal to $\Delta V/3$, where
$\Delta V$ is the variation of the sample volume caused by the
structural transition. This is also true for a structural
transition of the martensitic type if the distribution of the
martensitic variants that form upon the austenite -- martensite
transformation induced by the magnetic field is isotropic.
However, even in polycrystalline samples of a ferromagnetic shape
memory the special features of the texture may lead to substantial
dilatometric effects of martensitic transformation (up to 0.2\%)
[100]. In the case of single crystals or highly textured
polycrystalline samples, the dilatometric effects can be much
stronger. The experimental data taken from Refs [101-105] suggest
that upon thermocycling through the martensitic transition
temperature the formation of martensitic variants whose
orientation is favorable with respect to the applied magnetic
field is predominant. Hence the variations of linear dimensions
caused by the shift in the martensitic transition temperature in
such materials may be much larger than simple transition
striction.

If the magnetization of the martensitic phase differs from that of
the austenitic phase, application a magnetic field shifts the
structural transition temperature, stabilizing the phase with
larger magnetization [106]. This effect can be used to obtain
giant magnetostrains within the temperature interval of the
martensitic transformation. Reports of studies aimed at developing
functional Ni$_{2+x+y}$Mn$_{1-x}$Ga$_{1-y}$-based materials in
which giant magnetostrains are obtained through the shift of
martensitic transition temperature can be found in Refs [107-111].
Dikshtein \textit{et al.} [108] observed a reversible shift of the
martensitic transition temperature for Ni$_{2+x}$Mn$_{1-x}$Ga $(x
= 0.16 - 0.19)$ alloys that was controlled by a magnetic field.
The shape memory effect induced by the magnetic field and the
related giant magnetically induced strains have been studied by
Cherechukin \textit{et al.} [109], who used polycrystalline
Ni$_{2+x-y}$Mn$_{1-x}$Fe$_y$Ga samples. The researchers found that
an admixture of iron improves the mechanical properties of
Ni$_{2+x}$Mn$_{1-x}$Ga alloys. Samples in the shape of plates were
trained for a two-way shape memory effect by thermocycling under a
load. The training resulted in an increase in the attainable
bending strain from 2\% for an untrained sample to 4.5\% for the
trained sample after multiple thermocycling. The shape memory
effect caused by a shift of the martensitic transition temperature
induced by a magnetic field was observed for the trained plate of
Ni$_{2.15}$Mn$_{0.81}$Fe$_{0.04}$Ga with $T_m \sim 313$~K. The
experiment proceeded as follows. The plate, which in the
martensitic state had a curved shape, was subjected to a magnetic
field $H = 100$~kOe at room temperature. The plate was then heated
in this field to 315~K and the temperature was stabilized. In
these conditions the bending strain amounted to roughly 3\%. When
the magnetic field was switched off, the plate passed into the
austenitic state and became unbent. Thus, the bending strain
$\Delta e = 3$\% had been induced by the magnetic field $H =
100$~kOe.

\begin{figure}[t]
\includegraphics[width=6cm]{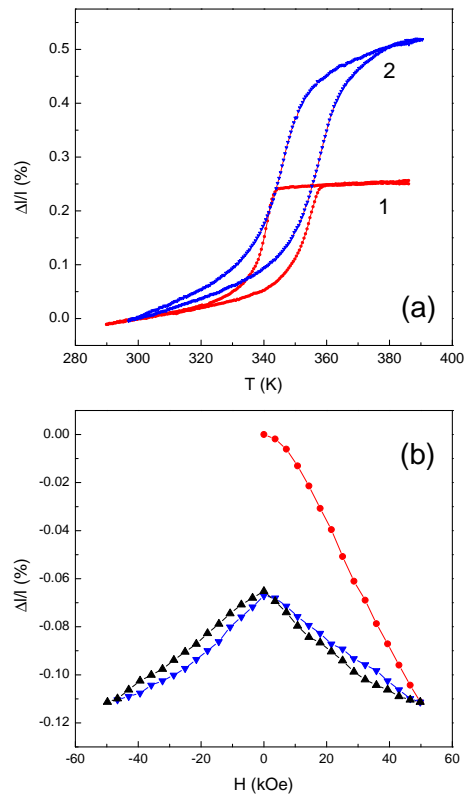}
\caption{(a) Temperature dependencies of the relative elongation
of a Ni$_{2.18}$Mn$_{0.82}$Ga sample prepared by the powder
metallurgy method without precompression (1) and after
precompression of 2\% in the martensitic state (2); (b)
magnetically induced strain (measured at $T = 351$~K) in a
Ni$_{2.18}$Mn$_{0.82}$Ga sample with a two-way shape memory
effect.}
\end{figure}

The type of process by which Ni$_{2+x}$Mn$_{1-x}$Ga alloys are
prepared strongly affects the magnetomechanical properties of
these alloys. One method used in powder metallurgy amounts to
sintering the fine powder by an electrical discharge under
pressure. The reversible change in the linear dimensions of such
samples upon a martensitic transformation reaches 0.2\%. Giant
magnetostrains in Ni$_{2+x}$Mn$_{1-x}$Ga $(x = 0.16 - 0.20)$
alloys fabricated by this method have been attained only after
preliminary training [111, 112]. The training of a
Ni$_{2.18}$Mn$_{0.82}$Ga sample amounted to the following. The
sample in the martensitic state was subjected to a uniaxial
compression of up to 2\%. After the sample was unloaded, the
residual strain amounted to 1.1\%. Upon the first heating -
cooling cycle, the sample demonstrated a shape memory effect and
almost completely restored its initial shape. Subsequent heating -
cooling cycles revealed the presence of a two-way shape memory.
Figure~12a shows that the multiple reversible change in the length
of this sample in the martensitic transition amounted to 0.5\%.
Measurements of magnetostrains in the given sample were conducted
at a temperature fixed in the interval of direct martensitic
transformation. It was found that the increase in the dilatometric
effect of the martensitic transformation in the sample with
two-way shape memory leads to a rise in the value of magnetically
induced strains. In the untrained Ni$_{2.18}$Mn$_{0.82}$Ga sample
the magnetically induced strain measured in the temperature
interval of the martensitic transition did not exceed 0.02\%,
while in the sample with two-way shape memory the values of
magnetically induced strain under the same conditions were as high
as $\approx 0.12$\%. However, such a strain, as shown in Fig.~12b,
was observed only during the first on - off magnetic field cycle,
while upon subsequent cycles the reversible value of magnetically
induced strain amounted to 0.05\%.

These results suggest that the value of magnetically induced
strains caused by the shift in the martensitic transition
temperature brought on by the magnetic field directly depends on
the dilatometric effect of the transformation. Although obtaining
large magnetically induced strains through the shift of the
martensitic transition temperature requires high magnetic fields,
this method has certain potential since the magnetically induced
strains in this case are comparable to the striction of the phase
transitions induced by a magnetic field in the following alloys,
which are being actively investigated: Fe-Rh [113], Mn-As [114],
La(Fe$_x$Si$_{1-x}$)$_{13}$ [115], and
Gd$_5$(Si$_x$Ge$_{1-x}$)$_4$ [116]. Among these materials,
La(Fe$_x$Si$_{1-x}$)$_{13}$ exhibits the highest value of linear
magnetostriction, $\Delta L/L \approx 0.3$\% in a 70~kOe magnetic
field [115]. The advantage of the Ni$_{2+x}$Mn$_{1-x}$Ga alloys is
the possibility of attaining giant magnetically induced strains at
temperatures much higher than room temperature, which is important
in some applications.

\subsection{Magnetostrain caused by the reorientation
of martensitic variants}

Shape memory ferromagnetic alloys are unique in the sense that
they make it possible to attain giant magnetically induced strains
caused by the reorientation of the martensitic variants by a
magnetic field. This mechanism was first proposed by Ullakko [117]
and was realized by Ullakko \textit{et al.} [102], who observed a
0.2\% strain induced by a 8~kOe magnetic field in a Ni$_2$MnGa
single crystal of nonstoichiometric composition with a martensitic
transition temperature $T_m \approx 276$~K. This was followed by
the paper of Tickle \textit{et al.} [118], who used this mechanism
to achieve irreversible (1.3\%) and reversible cyclic (0.5\%)
strains induced by a magnetic field $H < 10$~kOe in a
Ni$_{51.3}$Mn$_{24.0}$Ga$_{24.7}$ single crystal. Finally, in the
year 2000, Heczko \textit{et al.} [74] and Murray \textit{et al.}
[119] reported a 6\% magnetic field induced strain in the
martensitic phase of single crystalline samples of
nonstoichiometric compositions.

\begin{figure}[t]
\includegraphics[width=7cm]{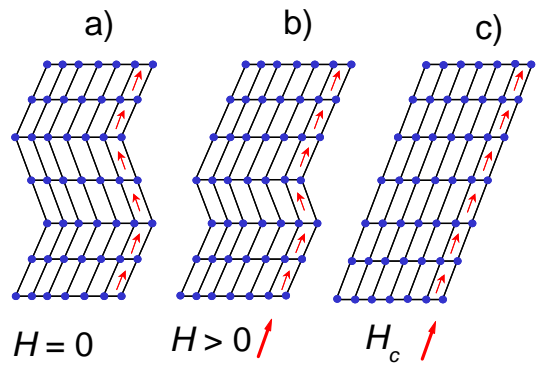}
\caption{Redistribution of martensitic variants in a magnetic
field. Variants that are favorably oriented with respect to the
applied magnetic field grow at the expense of unfavorably oriented
variants [120].}
\end{figure}

The mechanism by which a magnetic field transforms the magnetic
domains is illustrated schematically in Fig.~13 [120]. As is
known, the cooling of shape memory alloys to temperatures below
$M_f$ leads to the formation of self-accommodated martensitic
variants (Fig.~13a). In this representation, the boundaries of the
ferromagnetic domains coincide with those of the martensitic
variants. Under certain conditions a magnetic field \textbf{H} can
lead to growth of martensitic variants whose magnetic moment is
oriented favorably with respect to the magnetic field (Fig.~13b).
This process changes the shape of the sample. Ideally, at a
certain critical value of the magnetic field, $H_C$, all
martensitic variants align themselves along the direction of the
magnetic field (Fig.~13c). Such behavior is corroborated by the
results of direct optical observations done by Chopra \textit{et
al.} [121] on a Ni$_{53.8}$Mn$_{23.7}$Ga$_{22.5}$ single crystal.
Indeed, when a magnetic field is applied to the sample, the volume
fraction of favorably oriented martensitic variants was found to
increase at the expense of the volume fraction of the unfavorably
oriented martensitic variants.

The maximum value of the magnetically induced strain caused by
redistribution of martensitic variants is determined by the
intrinsic strain in the initial cubic lattice in the martensitic
transition, $1 - c/a$. For instance, for the system of
Ni$_{2+x+y}$Mn$_{1-x}$Ga$_{1-y}$ alloys the typical value $c/a
\approx 0.94$ makes it possible to attain strains of the order of
6\%. Magnetic fields needed to achieve such strains are roughly
10~kOe. The giant values of the magnetically induced strain
observed in these alloys are caused not only by the large
distortions of the initial phase in the martensitic transition but
also by the favorable ratio of the effective elastic modulus that
blocks the motion of the martensitic domain walls to the uniaxial
magnetocrystalline anisotropy constant $K_u$.

To observe the giant strains induced by magnetic field in
Ni$_{2+x+y}$Mn$_{1-x}$Ga$_{1-y}$ alloys, samples are usually
placed in a biasing magnetic field or are subjected, prior to
experiments, to uniaxial stresses. The same effect can be achieved
by changing the orientation of the magnetic field. Moreover, there
have been reports about cyclic reproducible magnetically induced
strains observed in unstressed samples.

The typical values of magnetically induced strains observed in
experiments with unstressed samples are 0.2--2\% [68, 70, 94, 102,
104, 122-128]. The highest magnetically induced strains have been
observed within temperature ranges from $M_s$ to $M_f$ and from
$A_s$ to $A_f$; outside these ranges magnetically induced strains
are negligible [104, 123]. This suggests that in this case the
leading role in generating magnetically induced strains is played
by the mobility of the boundaries of the martensitic variants. In
addition to observations of magnetically induced strains within
the temperature range of the martensitic transition, there have
been reports about large ($\sim 5$\%) reversible variations of the
linear dimensions of samples upon thermocycling through the
temperatures of martensitic [103, 104] and intermartensitic [70]
transitions in the presence of a magnetic field. Such a two-way
shape memory effect observed in the presence of a magnetic field
can be explained by the predominant growth of martensitic variants
that are favorably located with respect to the magnetic field.

The magnetically induced strains observed in experiments involving
prestressed samples or samples placed in a biasing magnetic field
or changes in the orientation of the magnetic field may be as high
as 6\% when $T < M_f$ [58, 74, 75, 81, 83, 118, 119, 120-140].
Such high values of magnetically induced strains can be explained
by the fact that uniaxial compression [119] or the application of
a biasing field that is perpendicular to the operating magnetic
field [118] facilitates the formation of a single-variant
martensitic state. A sample prepared in this way consists of
practically one martensitic variant (or, perhaps, several
equivalent variants) with the easy magnetization axis directed
along the vector of the biasing magnetic field. The application of
a magnetic field in the perpendicular direction leads to growth of
the martensitic variants whose easy magnetization axes are
directed along the operating magnetic field. This occurs if the
magnetocrystalline anisotropy energy density exceeds the effective
elastic modulus of the displacement of martensitic domain
boundaries. Such a process leads to magnetically induced strains
of the order of $1 - c/a$. When the operating magnetic field is
switched off in the absence of an external load or a biasing
magnetic field, the acquired magnetostrains are preserved.
Reversible magnetically induced strain is achieved in the presence
of an external load [119, 129] or a biasing magnetic field [118,
122]. Since the magnetically induced strain caused by the
transformation of the martensitic variants is observed in weak
magnetic fields, it can easily be suppressed by external loads, as
shown in Fig.~14.

\begin{figure}[t]
\includegraphics[width=7cm]{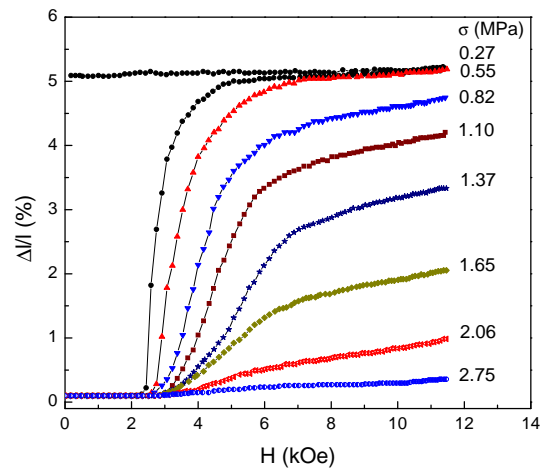}
\caption{Magnetically induced strains in a
Ni$_{49.8}$Mn$_{28.5}$Ga$_{21.7}$ single crystal for different
values of an external uniaxial load [119].}
\end{figure}

A comparison of the experimental data for different single
crystals shows that the effect of redistribution of martensitic
variants depends on many factors. The mobility of martensitic
domain walls depends on the method by which the sample was
prepared and on the thermal treatment of the sample. For instance,
Tickle \textit{et al.} [118] reported on a reversible magnetically
induced strain of 0.5\% in single crystal
Ni$_{51.3}$Mn$_{24.0}$Ga$_{24.7}$ ($M_s = 263$~K), and Murray
\textit{et al.} [119] observed a reversible magnetically induced
strain of 6\% in single crystal Ni$_{49.8}$Mn$_{28.5}$Ga$_{21.7}$
($M_s = 318$~K), although the preliminary training that the
samples underwent so that a single-variant state could be achieved
was almost the same (an external load of about 1~MPa). It is also
interesting that the composition of the single crystal used by
Murray \textit{et al.} [119] varied along the direction of growth
and that the 6\% magnetically induced strain was observed in a
sample cut from the beginning of the crystal; for samples cut from
the middle of the single crystals no magnetically induced strain
was observed [141].

\begin{table*}[t]
\caption{Magnetically induced strains in
Ni$_{2+x+y}$Mn$_{1-x}$Ga$_{1-y}$ alloys.}
\begin{tabular}{|c|c|c|c|c|}

\hline

Composition (at.\%)  &  $M_s$ (K) & MFIS (\%) & $K_u$ ($10^6$
erg/cm$^3$)  &  References \\

\hline

Unspecified & $\sim 276$ & 0.2 & 1.2  &  [102] \\

Ni$_{51.3}$Mn$_{24}$Ga$_{24.7}$ & $\approx 264$ & 0.5 - 1.3 & 2.45
& [118, 122] \\

Ni$_{52}$Mn$_{22.2}$Ga$_{25.8}$ & 289 &  0.3 &  & [123] \\

Ni$_{49.5}$Mn$_{25.4}$Ga$_{25.1}$ & $\sim 175$ & 0.4 & & [124] \\

Ni$_{48}$Mn$_{31}$Ga$_{21}$ & 301 & 5.1 & 1.7 &  [74, 135] \\

Ni$_{48}$Mn$_{30}$Ga$_{22}$ & 308 & 5.0 & & [75] \\

Ni$_{49.8}$Mn$_{28.5}$Ga$_{21.7}$ & 318 & 6.0 & 1.5 & [119] \\

Ni$_{47.4}$Mn$_{32.1}$Ga$_{20.5}$ & & 5.7\footnote{Shear strain} &
& [130] \\

Ni$_{52.3}$Mn$_{23.1}$Ga$_{24.6}$ & 251 & 0.12 & & [94] \\

Ni$_{52}$Mn$_{24.4}$Ga$_{23.6}$ & 276 & 0.6 - 1.2 & & [70, 104] \\

Ni$_{52}$Mn$_{23}$Ga$_{25}$ & $\approx 310$ & 0.27 & & [125] \\

Ni$_{49.6}$Mn$_{28.4}$Ga$_{22}$ & $\approx 306$ & 0.5 - 5.0 & 1.86
& [81, 131, 132, 134] \\

Ni$_{49.7}$Mn$_{28.7}$Ga$_{21.6}$ & $\approx 305$ & 5.0 - 5.3 &
2.48 & [81, 131, 132] \\

Ni$_{46.6}$Mn$_{29.5}$Ga$_{23.9}$ & $> 300$ & 2.2 & & [235] \\

Ni$_{48.1}$Mn$_{29.4}$Ga$_{22.5}$ & 270 & 0.3 & 1.5 & [81] \\

Ni$_{48.2}$Mn$_{30.8}$Ga$_{21}$ & 307 & 7.3 & 2.13 & [81] \\

Ni$_{48.7}$Mn$_{30.1}$Ga$_{21.3}$ & 302 & 4.5 & & [133] \\

Ni$_{53}$Mn$_{22}$Ga$_{25}$ & $\sim 305$ & 1.8 & & [68] \\

Ni$_{48.8}$Mn$_{29.7}$Ga$_{21.5}$ & 337 & 9.5\footnote{
Irreversible strain in the seven-layered martensitic phase} & &
[59] \\

Ni$_{49}$Mn$_{29.6}$Ga$_{21.4}$ & 306 & 3.8 & & [136, 137] \\

Ni$_{48.8}$Mn$_{28.6}$Ga$_{22.6}$ & $\approx 300$ & 5.0 & 2.0 &
[83] \\

Ni$_{48.5}$Mn$_{30.3}$Ga$_{21.2}$ & 297 & 6.0 & 1.7 & [58] \\

Ni$_{49.7}$Mn$_{29}$Ga$_{21.3}$ & $> 310$ & 2.6\footnote{At a
frequency of 2~Hz under an external load of 1.86~MPa} & $\approx
1.8$ & [140] \\

\hline

\end{tabular}
\end{table*}

The existing data on magnetically induced strains (MFIS) in single
crystalline Ni$_{2+x+y}$Mn$_{1-x}$Ga$_{1-y}$ caused by the
redistribution of the martensitic variants are systematized in
Table~1. These data suggest that giant magnetically induced
strains are observed as in alloys with slight deviations from
stoichiometry, as in alloys with a considerable excess of
manganese. The largest values of magnetically induced strains have
been achieved in samples with Mn content higher than 28 but lower
than 31 at.\% [58, 74, 75, 81, 83, 119, 131-140].

Shanina \textit{et al.} [81] assumed that the free electron
concentration plays an important role in achieving giant
magnetically induced strains in  Ni$_{2+x+y}$Mn$_{1-x}$Ga$_{1-y}$
alloys. Studies of non-stoichiometric alloys with different values
of magnetically induced strain revealed that alloys with a high
concentration of free electrons exhibit the largest magnetically
induced strain. Such alloys have the most pronounced metallic
nature of the interatomic bonds.

\section{Shape memory intermetallic compounds}

In addition to the Heusler alloys Ni$_{2+x+y}$Mn$_{1-x}$Ga$_{1-y}$
discussed above, structural phase transformations of the
martensitic type in the ferromagnetic matrix occur in several
other Heusler alloys, in ternary intermetallic compounds Co-Ni-Al
and Co-Ni-Ga, and in iron-based alloys such as Fe-Pd, Fe-Pt, and
Fe-Ni-Co-Ti. In this section we will also mention some
ferromagnets in which the martensitic transition is of a
nonthermoelastic nature.

The giant magnetically induced strains achieved in
Ni$_{2+x+y}$Mn$_{1-x}$Ga$_{1-y}$ alloys have stimulated the search
for, and intensive studies of, shape memory ferromagnets based on
other representatives of Heusler alloys. In this respect the alloy
that is being most actively investigated is Ni$_2$MnAl. As for
Cu$_2$MnAl-based Heusler alloys, they retain their ferromagnetic
properties only when deviations from stoichiometric composition
are small. Although a substantial decrease in Mn content in these
alloys does lead to the emergence of a martensitic transition, it
is accompanied by degradation of the ferromagnetic properties, so
that the compounds begin to resemble a superparamagnet or spin
glass.

Many intermetallic compounds experience ferromagnetic and
structural transitions of the martensitic type. In some of these
compounds, however, the type of martensitic transition
(thermoelastic or nonthermoelastic) has yet to be established. In
many compounds the structural transformation temperature $T_m$
exceeds the Curie temperature $T_C$, which makes it impossible to
attain giant magnetically induced strains via the shift of the
martensitic transition temperature. The processes of
redistribution of martensitic variants in the compounds described
in the sections below have never been studied, which makes it
possible to think of these compounds as materials in which
magnetic control of the shape memory effect is possible.

\subsection{Fe--Ni--Co--Ti}

The low-temperature body-centered tetragonal $\alpha$-phase in
iron and its alloys is ferromagnetic. Hence for iron alloys one of
the necessary conditions for strain caused by the redistribution
of martensitic variants in a magnetic field to appear is met. If
such elements as nickel, manganese, ruthenium, carbon, or nitrogen
are added to iron, the temperature of the polymorphic
transformation lowers, and with a sufficiently high content of
doping elements this transformation proceeds according to the
martensitic mechanism. It would seem that iron alloys are ideal
materials for attaining giant strains induced by a magnetic field.
This is not the case, however: neither a pronounced shape memory
effect nor a giant magnetically induced strain has been observed
in iron alloys. The reason is that the martensitic transformation
in iron alloys is not thermoelastic. The high-temperature phase
has a low elastic limit. The Bain strain, which describes the
transformation of the high-temperature fcc lattice into the
low-temperature body-centered tetragonal lattice, incorporates
compression of 17\% along the cubic axis of the high- temperature
phase and elongations of roughly 12\% in the directions
perpendicular to this axis. The symmetry of the low-temperature
phase allows for the formation of several crystallographic
variants of the high-temperature phase in the reverse
transformation.

The situation can be improved by narrowing the gap between the
lattices of the austenitic and martensitic phases and by raising
the elastic limit of both phases. To achieve this, the alloy
should be doped in such a way so as to cause particles that are
coherently coupled to the matrix to precipitate in the
high-temperature phase. After the martensitic transformation is
completed, the precipitate is inherited by the martensite, and the
low-temperature phase experiences tetragonal distortions. As a
result, the structures of the initial and martensitic phases
resemble each other more closely and the intrinsic strain in the
transformation diminishes. The presence of a precipitate
strengthens both the austenitic phase and the martensitic phase
and, as a result, hampers the process of generation of plastic
strain in the transformation. This idea was expressed by Kokorin
and Chuistov [142] and was soon corroborated by experiments: a
thermoelastic martensitic transformation was realized in
Fe$_{57}$Ni$_{23}$Co$_{10}$Ti$_{10}$ after aging of the
low-temperature martensitic phase of this alloy [143].

Kakeshita \textit{et al.} [144] reported their observations of a
martensitic transformation induced by a magnetic field in an
Fe$_{43}$Ni$_{33}$Co$_{10}$Ti$_4$ alloy (weight fractions) with
$M_s = 127$~K. Since the magnetic field was switched on when $T >
M_s$, the rise in the martensitic transformation temperature is
caused by the difference in the magnetizations of the austenitic
and martensitic phases. The martensitic transformation induced by
the magnetic field may be accompanied by changes in the linear
dimensions of the sample, but such a measurement was not performed
in Ref.~[144].

\subsection{Fe--Pd}

Fe--Pd alloys, which are close in composition to Fe$_3$Pd, can
undergo a thermoelastic martensitic transformation with small
hysteresis from the fcc phase to the face-centered tetragonal
phase with $c/a < 1$ [145-147]. The structural transformation is
preceded by a softening of the elastic modulus $C^{\prime}$ [148,
149]. The highest temperature at which the martensitic transition
occurs in the Fe--Pd system, $M_s$, is approximately 273~K and
rapidly decreases with increasing Pd content [147, 150]. Admixture
of Ni or Co also significantly lowers $M_s$ [150]. The
tetragonality parameter $c/a$ of the martensitic phase rapidly
increases with decreasing temperature [151] and in
Fe$_{70.3}$Pd$_{29.7}$ reaches the value of 0.91 at 153~K.

Systematic studies of the composition dependence of the Curie
temperature for Fe--Pd alloys have yet to be carried out. For the
Fe$_{70}$Pd$_{30}$ alloy, the Curie temperature $T_C = 573$~K
[152] and varies little with Pd content. At 333~K in the
austenitic phase of Fe$_{70}$Pd$_{30}$ the saturation
magnetization is roughly 1100~emu/cm$^3$ and the
magnetocrystalline anisotropy constant is $K_1 \sim -5 \times
10^3$ erg/cm$^3$ [152]. The fact that $K_1$ is negative implies
that the easy magnetization axis is directed along the $\langle
111 \rangle$ axis. The alteration of the crystal structure in the
process of martensitic transformation strongly affects the
magnetic properties of Fe--Pd alloys. The saturation magnetization
of the martensitic phase of Fe$_3$Pd is equal to 1400~emu/cm$^3$
and is weakly temperature-dependent [153, 154]. Measurements of
the uniaxial magnetocrystalline anisotropy constant of the
martensitic phase carried out on polycrystalline samples by Matsui
and Adachi [155] and Klemmer \textit{et al.} [156] revealed a
large value of $K_u$, roughly 10$^7$~erg/cm$^3$, which implies
that the coupling of the elastic and magnetic subsystems is
sufficiently strong for the emergence of magnetically induced
strains caused by the redistribution of the martensitic variants.
Note, however, that Cui and James [152] recently observed a much
smaller uniaxial magnetocrystalline anisotropy constant $K_u = 3.1
\times 10^5$ erg/cm$^3$ in an Fe$_{70}$Pd$_{30}$ single crystal.
There is also contradictory information about the direction of the
easy magnetization axis in the martensitic phase. For instance,
Matsui and Adachi [155] and Muto \textit{et al.} [157] reported
that the easy magnetization axis in the martensitic phase is
directed along [001]. On the other hand, there have been reports
(see Refs [152, 154]) of measurements involving single crystalline
samples converted to a single-domain state in which the [001] axis
was found to be the hard magnetization axis, while the easy
magnetization axis was found to be directed along [100]. These
contradictions are probably a reflection of the experimental
difficulties in determining the magnetic parameters of the
low-temperature phase, difficulties caused by the existence of
martensitic variants with different crystallographic orientations.

Studies of magnetically induced strains in the martensitic phase
of Fe$_{70}$Pd$_{30}$ have been conducted by James and Wuttig
[154]. Under the combined action of uniaxial stresses and the
operating and reference magnetic fields, a 0.6\% strain caused by
the redistribution of martensitic variants was reached in the
martensitic phase of a single-crystal sample.

Magnetically induced strain in an Fe$_{68.8}$Pd$_{31.2}$ single
crystal alloy with a martensitic transition temperature of 225~K
has been studied by Koeda \textit{et al.} [158]. The experiment
was conducted in the following way: first the sample was cooled to
liquid nitrogen temperature, and then it was subjected to a cyclic
magnetic field with a strength of up to 40~kOe (the direction of
the magnetic field was changed from positive to negative). In the
first on--off magnetic-field cycle the sample elongated by 3\%,
while in the subsequent cycles the reversible value of the
magnetically induced strain amounted to roughly 0.1\%.

Furuya \textit{et al.} [159] and Kubota \textit{et al.} [160]
studied the magnetically induced strain in ribbon samples of the
Fe$_{70.4}$Pd$_{29.6}$ prepared by quenching from melt. The grains
of the quenched samples were elongated along the direction of
rolling, and the variable parameters were the rate of quenching
from the liquid state and the annealing time; these parameters
determine the size distribution of the grains, the nature of the
grain boundaries, and the extent to which the grains are
texturized. A magnetic field of up to 10~kOe was applied
perpendicularly to the surface of the ribbon. The greatest strain
was observed in samples with grains whose (100) plane was parallel
to the sample surface. The strain induced by the magnetic field
increased with temperature up to 0.18\% (note, for the sake of
comparison, that the magnetostriction of the Terfenol-D,
Fe$_2$Tb$_{0.37}$Dy$_{0.23}$, in a 10~kOe field is 0.17\%).

\begin{figure}[b]
\includegraphics[width=7cm]{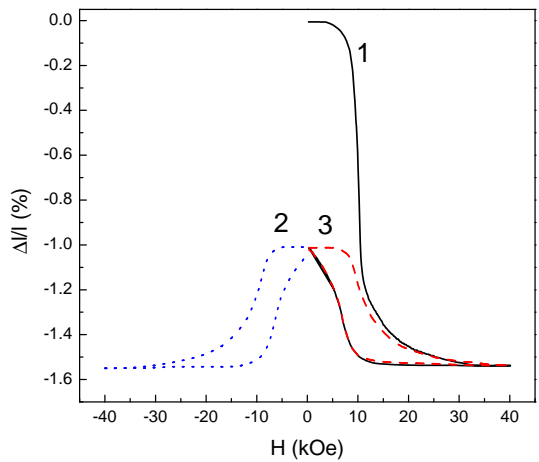}
\caption{Magnetically induced strains in a Fe$_3$Pt single crystal
in the first (1) and subsequent (2 and 3) on--off cycles of a
magnetic field \textbf{H} $\parallel$ [001] at $T = 4.2$~K [171].}
\end{figure}

\subsection{Fe--Pt}

In disordered Fe--Pt alloys, which are close in composition to the
stoichiometric alloy Fe$_3$Pt, the martensitic transformation from
the high-temperature fcc phase to the low-temperature
body-centered tetragonal phase proceeds with substantial (up to
450~K in the Fe$_{76}$Pt$_{24}$ alloy [161]) hysteresis. The
degree of shape restoration in the heating of a sample that had
been deformed in the martensitic state is small. Annealing leads
to the ordering of these alloys in a Cu$_3$Au (L1$_2$) structure.
Such ordering is accompanied by a rise in the Curie temperature
[162] and a drop in the martensitic transition temperature and
leads to a substantial decrease in hysteresis [161] and the
appearance of the shape memory effect [163-168].

As a result of atomic ordering, the martensitic transformation
changes from non-thermoelastic to thermoelastic. The transition
proceeds to the low-temperature face-centered tetragonal phase and
is preceded by a softening of the lattice of the high-temperature
phase as the martensitic transformation temperature is approached
[169]. The ordering makes the lattices of the initial and
martensitic phases resemble each other more closely, facilitates
elastic accommodation, and eliminates plastic relaxation of the
stresses generated in the process of formation of martensite
crystals in the initial phase.

The Curie temperature of Fe$_3$Pt is roughly 450~K, so that the
austenitic and martensitic phases in alloys close in composition
to the stoichiometric alloy are ferromagnetic [170]. Thus, the
conditions needed for observing magnetically induced strains
(thermoelastic martensitic transformation and the ferromagnetism
of the martensite) in ordered Fe--Pt alloys are met.

Magnetically induced strains in the single crystal Fe$_3$Pt, whose
martensitic transformation temperature is 97~K, have been studied
by Kakeshita \textit{et al.} [171]. The lattice constant $a$ of
the cubic L1$_2$ phase at 100~K is 3.73~\AA, and the lattice
parameters of the body-centered tetragonal martensitic phase at
77~K are $a = 3.77$~\AA ~and $c = 3.66$~\AA, so that $c/a = 0.97$.
The researchers did not determine the direction of the easy
magnetization axis -- it was assumed that this is the $c$ axis, as
in Ni$_2$MnGa.

The sample cooled to liquid helium temperatures was subjected to a
cyclic magnetic field with a strength of up to 40~kOe. As
Figure~15 shows, when the field was switched on for the first
time, the sample contracted by 1.5\% along the direction of the
field, and became 0.5\% longer when the field was switched off.
Subsequent on--off magnetic-field cycles revealed a reversible
change in the sample length of $\sim 0.5$\%, but hysteresis was
substantial (up to about 10~kOe).

\subsection{Ni$_2$MnAl}

The ordered high-temperature phase of the ternary system
Ni--Mn--Al undergoes a thermoelastic martensitic transformation
upon cooling [172-174]. The degree of ordering of the
high-temperature phase depends on thermal treatment: under
quenching from temperatures higher than 1000~$^{\circ}$C austenite
has a B2 structure which, however, changes to L2$_1$ structure
after ageing the samples at temperatures ranging from 350 to
400~$^{\circ}$C [174]. The crystal structure of the martensitic
phase and the martensitic transformation temperatures in
Ni--Mn--Al alloys depend on composition [172-175]. For instance,
for Ni$_{50+x}$Mn$_{25}$Al$_{25-x}$ alloys the temperature of the
onset of the martensitic transition, $M_s$, is in the interval
from 200 to 250~K for $0 \le x \le 3$ [176, 177].

Neutron diffraction studies at liquid helium temperatures have
shown [178] that Ni$_2$MnAl of stoichiometric composition has
antiferromagnetic ordering. The type of magnetic ordering in
Ni--Mn--Al alloys strongly depends on the chemical ordering of the
high-temperature phase, which changes under thermal treatment.
Quenched samples have the B2 structure of the austenitic phase,
and they exhibit either antiferromagnetism or a spin glass state.
Ageing the samples after quenching leads to the formation of the
L2$_1$ structure of the austenitic phase, which has a
ferromagnetic ordering with a Curie temperature $T_C \sim 330$~K
[176].

Fujita \textit{et al.} [179] used the capacitance dilatometer
method to study the strains induced by magnetic fields in
polycrystalline and single crystalline
Ni$_{53}$Mn$_{25}$Al$_{22}$. For polycrystalline samples of this
alloy with $M_f \sim 226$~K the value of the magnetostrain was
found to be temperature-dependent. The maximum strain of roughly
0.01\% was observed at 228~K. The researchers assumed that this
was due to the presence of a small amount of the austenitic phase
near $M_f$, which facilitates the movement of the boundaries of
the martensitic variants [104]. In a 70~kOe field, the
magnetically induced strain in the single crystal sample amounted
to 0.1\%.

\subsection{Co--Ni--Al, Co--Ni--Ga}

Enami and Nenno [180] established that the Ni--Al alloy with a B2
structure exhibits the shape memory effect. The influence of a
cobalt admixture on the shape memory effect in Ni--Al was studied
by Kainuma \textit{et al.} [181], who found that Co--Ni--Al alloys
may undergo a thermoelastic martensitic transition and that the
temperature of this transition decreases with increasing the Co
content. The researchers also found that austenite in these alloys
is actually a two-phase solid solution, and because of this the
polycrystalline Co--Ni--Al alloys exhibit good plastic properties.

In addition to the Co--Ni--Al alloys, ternary Co--Ni--Ga alloys
have been studied. It was found [182-184] that these alloys also
exhibit shape memory and, in the case of a two-phase
microstructure, have good plastic properties. Shape memory in
Co$_{45}$Ni$_{25}$Ga$_{30}$ and Co$_{40}$Ni$_{33}$Al$_{27}$ was
studied by the three-point bending method involving lamellar
samples whose austenitic phase had a two-phase microstructure. It
was found that the initial shape of the samples was not restored
after the stresses were removed, which, probably, is due to the
presence of residual "inclusion phases".

The field and temperature dependencies of the magnetization of
Co--Ni--Al and Co--Ni--Ga exhibit a behavior typical of
ferromagnetic materials. The dependence of the martensitic and
ferromagnetic transition temperatures on composition in Co--Ni--Al
and Co--Ni--Ga has been studied in Refs [182-185]. In Co--Ni--Al
with fixed Al content (at 30 at.\%), the martensitic transition
temperature $T_m$ increases and the Curie temperature $T_C$
decreases as Ni content grows from 30 to 45 at.\%. An important
feature of the phase diagram built by the researchers is that, as
in the case of Ni$_{2+x}$Mn$_{1-x}$Ga alloys, $T_m$ and $T_C$
merge at $\sim 250$~K when Ni content reaches $\sim 35$~at.\%. A
further increase in Ni content makes $T_m$ higher than $T_C$ and
the martensitic transition takes place in the paramagnetic phase.
Such concentration dependencies of $T_m$ and $T_C$ were also found
to exist for Co--Ni--Ga alloys. For instance, with Ga content
fixed (at 30 at.\%), the martensitic transition temperature $T_m$
increases and the Curie temperature $T_C$ decreases as Ni content
grows from 18 to 33 at.\%. For the Co$_{47}$Ni$_{23}$Ga$_{30}$
alloy, the temperatures $T_m$ and $T_C$ merge at 370~K.

Morito \textit{et al.} [186] found that the value of the
magnetocrystalline anisotropy of the martensitic phase of
Co--Ni--Al, when measured in single crystal
Co$_{37}$Ni$_{34}$Al$_{29}$, amounts to $3.9 \times 10^7$
erg/cm$^3$. This value is comparable to the magnetocrystalline
anisotropy constants of Ni$_{2+x+y}$Mn$_{1-x}$Ga$_{1-y}$ alloys
[81]. Measurements of magnetically induced strain caused by the
redistribution of martensitic variants have shown, however, that
the reversible magnetically induced strains in this intermetallic
compound do not exceed 0.06\%, which is smaller than the values
for Ni$_{2+x+y}$Mn$_{1-x}$Ga$_{1-y}$ alloys by a factor of 100.
Morito \textit{et al.} [186] assumed that this is because the
sample was not in a single-variant state and hence only a fraction
of the maximum attainable magnetically induced strain was
detected. The crystal structure of martensite in
Co$_{37}$Ni$_{34}$Al$_{29}$ (L1$_0$, $c/a = 0.816$) also differs
significantly from that in Ni$_{2+x+y}$Mn$_{1-x}$Ga$_{1-y}$
alloys, a factor that could affect the mobility of the martensitic
variants.

\subsection{Co$_2$NbSn}

The ferromagnetic Heusler alloy Co$_2$NbSn undergoes a martensitic
transformation when it is cooled. The temperature of this
martensitic transition, $T_m = 253$~K, is much higher than the
ferromagnetic ordering temperature $T_C = 119$~K [187]. In the
low-temperature phase this compound has an orthorhombic crystal
lattice with parameters $a = 4.436$~\AA, $b = 4.397$~\AA, and $c =
5.939$~\AA ~[69, 188]. The magnetic moment in Co$_2$NbSn is on the
cobalt atoms and amounts to approximately $0.36 \mu _\mathrm{B}$
at 4~K [189]. Studies of the magnetic properties of the
martensitic phase of Co$_2$NbSn have shown (see Refs [188, 190,
191]) that magnetization does not reach saturation in fields up to
5~T. Such measurements conducted with the
Co$_{50.4}$Nb$_{25}$Sn$_{12.9}$Ga$_{11.7}$ alloy which does not
undergo a martensitic transformation, have shown that
magnetization becomes saturated in fields of approximately 0.5~T.
Ohtoyo \textit{et al.} [188] assumed, on the basis of these
measurements, that large magnetocrystalline anisotropy is a
characteristic feature of the martensitic phase of the Co$_2$NbSn
alloys. Variations in composition [188] and doping of Co$_2$NbSn
with Ni [69], V, Ga, and Al [192] do not lead to a rise in the
Curie temperature $T_C$. However, the temperature of the
martensitic transition decreases. For instance, in the
Co$_{50}$Nb$_{25}$Sn$_{20}$Ga$_5$ alloy, $T_m = 43$~K but $T_C =
101$~K [192].

\section{Theory of phase transitions
in cubic ferromagnets}

The experimental data suggest that both structural and magnetic
phase transformations occur in ferromagnetic alloys with shape
memory. Moreover, within certain intervals of structural and
magnetic parameters of ferromagnets these transformations overlap.
Intensive experimental studies of shape memory ferromagnets have
stimulated the development of the theory of coupled structural and
magnetic transformations.

Theoretical investigations of the interaction of the structural
and magnetic order parameters in various crystals have been
carried out by many researchers. The results of earlier work of
this kind have been gathered and generalized in Izyumov and
Syromyatnikov's monograph [193]. Many theorists have also examined
the transition from the cubic phase to the tetragonal in various
crystals. Landau theory of phase transitions has been used to
describe the given transition. This theory was used to study
transitions only in the elastic subsystem [193-198] and
transitions in systems where the structural and magnetic order
parameters interact [22, 199]. As applied to phase transitions in
Ni$_{2+x}$Mn$_{1-x}$Ga alloys, Landau theory has been developed in
Refs [66, 67, 91, 200-215].

\subsection{Total energy}

The sequence of phase transformations in Ni$_{2+x}$Mn$_{1-x}$Ga
alloys observed in experiments can be described by Landau theory
for coupled structural and magnetic phase transitions. The special
features of the crystal and magnetic structures of the
ferromagnetic Heusler alloys Ni--Mn--Ga require the introduction
of three interacting order parameters describing the variation of
the structure of the crystal lattice, its modulation, and the
magnetization. Within this system of parameters, both the
ferromagnetic cubic state and the ferromagnetic tetragonal state
(both modulated and unmodulated) can exist at low temperatures.

The distortions of the cubic lattice upon structural transitions
are described by homogeneous strains written in the form of linear
combinations of the components $e_{ii}$ of the strain tensor
[216]:

\begin{eqnarray}
 e_1&=&\frac{1}{3}(e_{xx} + e_{yy} + e_{zz}),\notag \\
e_2&=&\frac{1}{\sqrt{2}}(e_{xx} - e_{yy}), \\
e_3&=&\frac{1}{\sqrt{6}}(2e_{zz} - e_{xx} - e_{yy}).\notag
\end{eqnarray}

\noindent The strain $e_1$, which reflects changes in volume, does
not violate lattice symmetry. Symmetry breaking occurs because of
the strains $e_2$ and $e_3$, responsible for the transition of the
lattice from the cubic phase to the tetragonal. The given
transition is accompanied by the softening of the elastic moduli
combination $C_{11} - C_{12}$. The complete expression for the
free energy density, which describes the transition to the
tetragonal state, must contain the strains $e_4 = e_{xy}$, $e_5 =
e_{yz}$, and $e_6 = e_{zx}$, which lead to the distortion of the
unit cell to a symmetry lower than the tetragonal.

In describing the modulated state in the crystal brought on by the
presence of acoustic phonon modes of the type $(\frac{1}{3}
\frac{1}{3} 0)$, we must bear in mind that there are six different
directions of the modulation wave vector. They can be written as
follows: $\mathbf{k}_1 = \zeta (110)$, $\mathbf{k}_2 = \zeta
(011)$, $\mathbf{k}_3 = \zeta (101)$, $\mathbf{k}_4 = \zeta (1\bar
10)$, $\mathbf{k}_5 = \zeta (01\bar 1)$, and $\mathbf{k}_6 = \zeta
(\bar 110)$, where $\zeta = 1/3$. In view of this, there should,
as a rule, exist an order parameter containing twelve components
(six amplitudes and six phases): $\psi _1, \ldots, \psi_6$ and
$\varphi _1, \ldots, \varphi _6$, where $\psi _j = |\psi _j|
\text{exp}(\mathrm{i}\varphi _j)$. The atomic displacements
corresponding to each of these order parameters have the form
$\mathbf{u}_j(\mathbf{r}) = |\psi _j|\mathbf{p}_j\,
\text{sin}(\mathbf{k}_j\mathbf{r} + \varphi _j)$, where
$\mathbf{p}_1,\ldots,\mathbf{p}_6$ are the unit polarization
vectors directed along the axes $[\bar 110]$, $[0\bar 11]$,
$[10\bar 1]$, [110], [011], and [101], respectively.

The expression for the free energy density, which describes
structural transitions from the cubic phase, must be invariant
under spatial transformations of the point symmetry group $O_h$.
It consists of terms of three types:

\begin{equation}
F = F_e(e_j) + F_{\psi}(\psi _i) + F_{e\psi}(e_j,\psi _i).
\end{equation}

\noindent Here $F_e(e_j)$ is the elastic energy density, which
contains terms responsible for the anharmonicity of the elastic
subsystem with respect to the order parameter ($e_2, e_3$). The
expression for it has the form [216]

\begin{widetext}
\begin{equation}
F_e(e_j) = \frac{1}{2}(C_{11}+2C_{12})e^2_1 +
\frac{1}{2}a(e^2_2+e^2_3) + \frac{1}{2}C_{44}(e^2_4+e^2_5+e^2_6) +
\frac{1}{3}be_3(e^2_3-3e^2_2) + \frac{1}{4}c(e^2_2+e^2_3)^2,
\end{equation}
\end{widetext}

\noindent where the coefficients $a$, $b$, and $c$ are linear
combinations of the components of the elastic moduli of the
second, third, and fourth orders, respectively:

\begin{eqnarray}
a&=&C_{11} - C_{12},\notag \\
b&=&\frac{1}{6\sqrt{6}}(C_{111}-3C_{112}+2C_{123}),\\
c&=&\frac{1}{48}(C_{1111}+6C_{1112}-3C_{1122}-8C_{1123}).\notag
\end{eqnarray}

\noindent Since the right-hand side of Eqn~(3) contains terms of
the third order, the phase transition in the parameter ($e_2$,
$e_3$) is a first-order transition. As the point of the structural
transition to the tetragonal phase is approached, the elastic
modulus $a = C_{11} - C_{12}$ tends to zero, and near the
transition point ($T \to T_m$) it can be represented in the form
$a = a_0(T - T_m)$, where $T_m$ is the temperature of the
martensitic transition.

The complete expression for $F_{\psi}(\psi _j)$ can be found in
Refs [217, 218]. Here we consider the simplest version of
modulation, which allows for only one phonon mode
$\frac{1}{3}$(110). It is described by the order parameter $\psi =
|\psi|\text{exp}(\mathrm{i}\varphi)$ (to simplify the notation, we
have dropped the index $j$), which makes it possible to write the
expression for the density of the modulation part of the free
energy in the form

\begin{equation}
F_{\psi}(\psi) = \frac{1}{2}A|\psi|^2 + \frac{1}{4}B|\psi|^4 +
\frac{1}{6}C_0|\psi|^6 +\frac{1}{6}C_1\big[\psi ^6+(\psi
^{\ast})^6\big].
\end{equation}

\noindent The last term of this equation is minimized with respect
to the phase:

\begin{equation*}
\psi ^6 + (\psi ^{\ast})^6 =
|\psi|^6\big[\text{exp}(-\mathrm{i}6\varphi) +
\text{exp}(\mathrm{i}6\varphi)\big] = 2|\psi|^6\text{cos}6\varphi.
\end{equation*}

\noindent The minimum of the energy (5) occurs at $\varphi = \pm
\pi/6$, $\pm\pi/2$, $\pm 5\pi/6$ if $C_1 > 0$ and at $\varphi =
0$, $\pm\pi/3$, $\pm2\pi/3$, $\pi$ if $C_1 < 0$. In equation (5)
we introduce the quantity $C^{\prime} = C_0 - |C_1|$ and, for the
sake of stability, assume that $C^{\prime} > 0$. The parameter $A$
is temperature-dependent, and near the temperature of transition
to the modulated state ($T \to T_P$) it can be written in the form
$A = A_0(T - T_P)$ [20, 219, 220].

The part of the free energy density, $F_{e\psi}$, that connects
the strain, $e_i$, and the order parameter describing modulation
must be invariant under all symmetry operations related to $e_i$
and $\psi_j$. If we allow for the phonon mode $\frac{1}{3}$(110)
only, the expression for $F_{e\psi}$ has the form

\begin{equation}
F_{e\psi}(\psi,e_i) = \bigg(\frac{1}{\sqrt{3}}D_1e_1 +
\frac{2}{\sqrt{6}}D_2e_3 + D_3e_4\bigg)|\psi|^2.
\end{equation}

Equations (3), (5), and (6) completely determine the free energy
density of a cubic crystal, which makes it possible to describe
structural phase transitions from the cubic phase to phases of
lower symmetry.

Experiments have shown that structural transformations in
Ni$_{2+x}$Mn$_{1-x}$Ga alloys take place in the ferromagnetic
matrix, which makes it necessary to take into account the effect
that the magnetic subsystem has on the structural transitions. In
these alloys Mn atoms are the main carriers of magnetic moments,
both in the cubic and in the tetragonal phases [14]. This makes it
possible to describe Ni$_{2+x}$Mn$_{1-x}$Ga alloys using the model
of a single-sublattice magnetic subsystem with a macroscopic
magnetization vector \textbf{M}. In describing the behavior of the
magnetic subsystem we introduce, for the sake of convenience, the
dimensionless magnetization vector $\mathbf{m} = \mathbf{M}/M_0$,
where $M_0$ is the saturation magnetization.

The contribution of the magnetic subsystem to the total energy of
the cubic ferromagnet consists of two parts. The first is of
exchange origin and is needed to allow for the dependence of the
magnetization vector \textbf{m} on temperature. This term can be
written as follows:

\begin{equation}
F_{\text{ex}}(m) = \frac{1}{2}\alpha(m^2_x+m^2_y+m^2_z) +
\frac{1}{4}\delta_1(m^2_x+m^2_y+m^2_z)^2.
\end{equation}

\noindent Here $\alpha$ and $\delta_1$ are the exchange constants.
The parameter $\alpha$ of the exchange interaction is
temperature-dependent and near the Curie point, $T_C$, can be
written as $\alpha = \alpha_0(T - T_C)$. The second term is the
magnetic anisotropy energy of the cubic ferromagnet, which can be
written as

\begin{equation}
F_{\text{a}}(m_i) = K_1(m^2_xm^2_y+m^2_ym^2_z+m^2_zm^2_x),
\end{equation}

\noindent where $K_1$ is the first cubic anisotropy constant.

The expression for the free energy must also contain a terms that
relate the components of \textbf{M} with the other order
parameters of the system. The first of such terms connects the
magnetization components $m_i$ with the strains $e_i$ and has the
form

\begin{widetext}

\begin{equation}
F_{me}(m_i,e_j) = \frac{1}{\sqrt{3}}B_1e_1\mathbf{m}^2 +
B_2\bigg[\frac{1}{\sqrt{2}}e_2(m^2_x-m^2_y)+\frac{1}{6}e_3(3m^2_z-\mathbf{m}^2)\bigg]
+ B_3(e_4m_xm_y+e_5m_ym_z+e_6m_zm_x).
\end{equation}

\noindent This formula is the simplest expression for
magnetoelastic energy with phenomenological magnetoelastic
constants $B_1$, $B_2$, and $B_3$.

The second term describes the interaction between the
magnetization vector components $m_i$ and the order parameter
$\psi$ and can be written as follows:

\begin{equation}
F_{m\psi}(m_i,\psi) = \bigg[\frac{1}{3}N_1\mathbf{m}^2 +
N_2\bigg(m^2_z-\frac{1}{3}\mathbf{m}^2\bigg) +
N_3m_xm_y\bigg]|\psi|^2.
\end{equation}

\noindent Here the coefficients $N_i$ are the coupling parameters
between the magnetic and modulation subsystems.

As a result, we arrive at the following final expression for the
free energy density:

\begin{equation}
F = F_e(e_i) + F_{\psi}(|\psi|^2) + F_{e\psi}(e_i,|\psi|^2) +
F_{\text{ex}}(m) + F_{\text{a}}(m_i) + F_{me}(m_i,e_j) +
F_{m\psi}(m_i,|\psi|^2).
\end{equation}

This expression for the free energy density contains variables
that are not responsible for phase transitions, i.e., variables
that are indirect order parameters: $e_1$, $e_4$, $e_5$, and
$e_6$. After the free energy is minimized in all these variables,
some of the constants in Eqn~(11) become renormalized:

\begin{equation*}
B^{\prime}=B-2\bigg[\frac{D^2_1}{3(C_{11}+2C_{12})}+\frac{D^2_3}{C_{44}}\bigg],\quad
K=K_1-\frac{B^2_3}{2C_{44}},\quad
N^{\prime}_1=\frac{1}{3}N_1-\frac{D_1B_1}{6(C_{11}+2C_{12})},
\end{equation*}

\begin{equation*}
N^{\prime}_3=N_3-\frac{B_3D_3}{C_{44}},\quad
\delta=\delta_1-\frac{2B^2_1}{3(C_{11}+2C_{12})}.
\end{equation*}

After such renormalization has been carried out, the expression
for the free energy density of the cubic ferromagnet becomes

\begin{equation}
\begin{split}
F&=\frac{1}{2}a(e^2_2+e^2_3)+\frac{1}{3}be_3(e^2_3-3e^2_2)+\frac{1}{4}c(e^2_2+e^2_3)^2+
\frac{1}{2}A|\psi|^2+\frac{1}{4}B^{\prime}|\psi|^4+\frac{1}{6}C^{\prime}|\psi|^6+\frac{2}{\sqrt{6}}D_2e_3|\psi|^2+
\\
&+\frac{1}{2}\alpha\mathbf{m}^2+\frac{1}{4}\delta\mathbf{m}^4+K(m^2_xm^2_y+m^2_ym^2_z+m^2_zm^2_x)+
B_2\bigg[\frac{1}{\sqrt{2}}e_2(m^2_x-m^2_y)+\frac{1}{\sqrt{6}}e_3(3m^2_z-\mathbf{m}^2)\bigg]+\\
&+\bigg[N^{\prime}_1\mathbf{m}^2+N_2\bigg(m^2_z-\frac{1}{3}\mathbf{m}^2\bigg)+N^{\prime}_3m_xm_y\bigg]|\psi|^2.
\end{split}
\end{equation}
\end{widetext}

\subsection{Phase diagram at $T_m < T_C$}

In Ni$_{2+x}$Mn$_{1-x}$Ga alloys the temperatures of the
structural and magnetic phase transitions strongly depend on
composition. In view of this it would be interesting to study the
phase diagrams at $T_m < T_C$ (compositions close to
stoichiometric) and at $T_m \sim T_C$ (compositions with $x = 0.16
- 0.20$). The review of the experimental research in magnetic
anisotropy measurements (see Section~4.3) suggests that the first
cubic- anisotropy constant $K$ also depends on composition. For
instance, in alloys close to the stoichiometric composition, the
constant $K$ is positive [80], while in alloys with considerable
deviations from stoichiometry it proves to be negative [81]. In
view of this, when describing phase transitions in
Ni$_{2+x}$Mn$_{1-x}$Ga alloys by theoretical means, both cases, $K
> 0$ and $K < 0$, must be considered. It has turned out that the
second case leads to more complicated phase diagrams than the
first (see Refs [66, 67, 91, 200-210] and [211-215] for $K < 0$
and $K > 0$, respectively). In view of this we will focus on the
case where $K > 0$. The interested reader can find material on the
$K < 0$ case in the above-cited works [66, 67, 91, 200-210]. To be
more definite, we assume in what follows that the magnetostriction
constant $B_2$ is positive (this follows from the measurements of
magnetostriction constants [80]) and that the fourth-order elastic
modulus $c > 0$. But first we assume that there is no lattice
modulation, i.e., $|\psi| = 0$.

We begin with the case where the structural transition takes place
in a ferromagnetic matrix ($T_m < T_C$). Here in formula (12) we
can put $\mathbf{m}^2 = 1$ and go from the Cartesian coordinates
$m_x$, $m_y$, and $m_z$ to the polar $\theta$ and azimuthal $\phi$
angles of this vector. Minimization of the free energy (12) in
respect to $e_2$, $e_3$, $\theta$, and $\phi$ leads to the
following equilibrium states.

1. Cubic (C) and tetragonal (T) phases with magnetizations along
the [001] axis $(\theta = 0, \pi)$ and with strains determined
from the equations

\begin{equation}
e_2 = 0, \quad ae_3+be^2_3+ce^3_3+\frac{\sqrt{6}}{3}B_2 = 0,
\end{equation}

\noindent are stable at $b \le 0$ and exist in the region
specified by the inequality

\begin{equation}
a \ge
\frac{\sqrt{6}}{9}\frac{B_2c}{b}+\frac{2}{3}\sqrt{-bB_2\sqrt{6}}.
\end{equation}

\noindent At $b \ge 0$ the regions of existence of these phases
are separated by the branches of the discriminant curve

\begin{equation}
a^3 - \frac{b^2}{4c}a^2 - \frac{3\sqrt{6}}{2}abB_2 +
\frac{9}{2}cB^2_2 + \frac{\sqrt{6}}{3}\frac{b^3}{c}B_2 = 0
\end{equation}

\noindent of the cubic equation (13). Inside the region bounded by
the branches of the discriminant curve (15), both phases are
stable.

2. The orthorhombic (R) phase with magnetization along the [001]
axis ($\theta = 0, \pi$) and with strains determined from the
equations

\begin{equation}
\begin{split}
4cbe^2_3 &- 2b^2e_3 + ab +\frac{\sqrt{6}}{3}B_2 = 0,\\
e_2&=\pm\sqrt{-e^2_3-\frac{1}{c}+e_3\frac{2b}{c}},
\end{split}
\end{equation}

\noindent is stable at

\begin{equation}
\begin{split}
a\le\frac{b^2}{4c}-\frac{\sqrt{6}}{3}\frac{B_2c}{b}, \quad
&b<-\bigg(\frac{16\sqrt{6}}{9B_2c^2}\bigg)^{1/3},\\
a\le\frac{\sqrt{6}}{9}\frac{B_2c}{b}+&\frac{2}{3}\sqrt{-\sqrt{6}bB_2},\\
-\bigg(\frac{16\sqrt{6}}{9B_2c^2}\bigg)^{1/3}&<b\le0.
\end{split}
\end{equation}

For $b > 0$ there is a region in which a metastable phase
R$^{\prime}$ coinciding in symmetry with the phase R is stable,
while for $b < 0$ there is a region in which a metastable phase
C$^{\prime}$(T$^{\prime}$) coinciding in symmetry with the phase
C(T) is stable. The signs of the strains $e_3$ in the metastable
phases are opposite to those of the given strains in the phases R
and C(T). The existence regions of the metastable phases are
bounded by the following curves. At $b > 0$ (phase R$^{\prime}$),

\begin{equation}
\begin{split}
\sqrt{6}cK\frac{1-\sqrt{1-B^4_2/cK^3}}{3B_2}\le &b \le
\sqrt{6}cK\frac{1+\sqrt{1-B^4_2/cK^3}}{3B_2},\\
\frac{B^4_2}{cK^3}&\le 1,\\
a\ge-\frac{2K^2c}{3B^2_2}-\frac{B^4_2c}{9b^2K^2}&+\frac{\sqrt{6}bK}{3B_2}-\frac{B^2_2}{3K}-\frac{\sqrt{6}B_2c}{b},\\
a\le \frac{b^2}{4c}& - \frac{\sqrt{6}B_2c}{3b},
\end{split}
\end{equation}

and at $b < 0$ [phase C$^{\prime}$(T$^{\prime}$)],

\begin{equation*}
a \ge \frac{\sqrt{6}B_2c}{9b}-\frac{2\sqrt{-\sqrt{6}bB_2}}{3}
\end{equation*}

\noindent and the discriminant curve (15).

Note that according to Eqn~(1), the sign of the strain $e_3$
determines the sign of $c/a - 1$ (here $c$ and $a$ are the
structural lattice parameters). At $e_3 > 0$ the quantity $c/a -
1$ is positive, while at $e_3 < 0$ it is negative. If we assume
that the elastic moduli also depend on the composition of the
Ni$_{2+x}$Mn$_{1-x}$Ga alloys, equations (16) -- (18) suggest that
at certain values of $x$ stable and metastable phases of different
symmetries may coexist in these alloys. This fact agrees with the
experimental results of Inoue \textit{et al.} [221, 222], who
observed martensitic phases of different symmetries.

Symmetry considerations suggest that in addition to the above
states in the ferromagnet there can be similar phases with
magnetizations along the [100] and [010] axes, states whose
energies and regions of existence coincide with those given above.

Analysis of distortions of the initial cubic lattice in phases C
and T shows that these phases have the same tetragonal symmetry
and differ only in the values of spontaneous strains. In the C
phase these strains are determined by the distortions of the cubic
lattice caused by magnetostriction, while in the T phase they are
determined by structural distortions in the transition to the
martensitic state. The phase transition lines between the states
C, T, and R are determined by the condition that the phase
energies (12) be equal.

\begin{figure}[t]
\includegraphics[width=7cm]{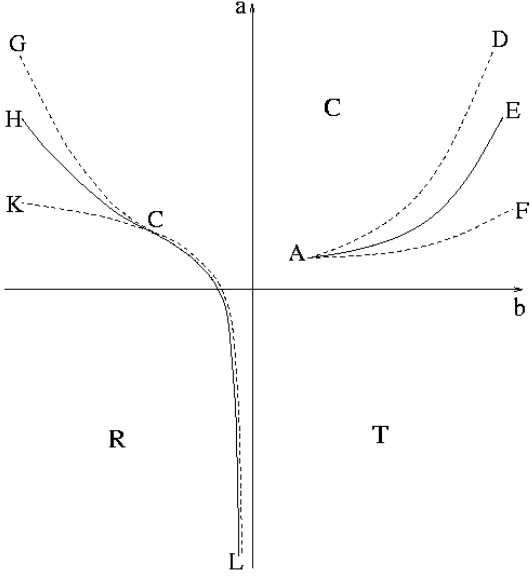}
\caption{Phase diagram of a cubic ferromagnet with $T_m < T_C$ in
the $(a, b)$ plane. C is the cubic phase with small tetragonal
distortions, T is the tetragonal phase, and R is the orthorhombic
phase. The magnetization in all phases is directed along the [001]
axis. The solid curves represent the phase transition lines and
the dashed curves the lines of loss of stability of the phases.}
\end{figure}

The phase diagram of the cubic ferromagnet in the $(a, b)$ plane
at $T_m < T_C$ is depicted in Fig.~16. Depending on the values of
the elastic moduli of the second $(a)$ and third $(b)$ orders, the
following structural transformations can occur in the ferromagnet.
At $b > 0$, a first-order phase transition from phase C to phase T
occurs on the line $AE$ determined by the equation

\begin{equation}
a = \frac{2}{9}\frac{b^2}{c} + \frac{\sqrt{6}B_2c}{b}.
\end{equation}

\noindent This transition is accompanied by a discontinuity of the
strain $e_3$ and is a martensitic transformation. In relation to
symmetry, it is an isostructural transition and ends at point $A$
(the critical point) with coordinates $(18cB^2_2)^{1/3}$ and
$(9\sqrt{6}B_2c^2)^{1/3}$. To the left of point $A$ the transition
from phase C to phase T proceeds smoothly, without discontinuity
in $e_3$. At $b \le 0$, a martensitic transformation (a
first-order structural phase transition) from the cubic phase C to
the orthorhombic phase R, accompanied by discontinuities in the
strains $e_2$ and $e_3$, occurs on the line $CH$. On the line $CL$
a second-order structural phase transition between these two
phases occurs. The expression for the line $CH$ of a first-order
phase transition can be found by equalizing the energies of phases
C and R. Line $CL$ of a second-order phase transition is
determined by using the equal sign in Eqn~(14). The critical point
$C$ where the first-order phase transition is terminated has the
coordinates

\begin{equation*}
\bigg(\frac{\sqrt{6}}{3}-\frac{1}{2}\bigg)(B^2_2c)^{1/3}, \quad
-\bigg(\frac{16\sqrt{6}}{9B_2c^2}\bigg)^{1/3}
\end{equation*}

\noindent Clearly, the coordinates of the points $A$ and $C$ are
determined by the size of the magnetostriction $B_2$. At $B_2 = 0$
the phase diagram coincides with that of a nonmagnetic cubic
crystal [193].

\subsection{Phase diagram at $T_m \approx T_C$}

Now let us examine the phase diagram of the cubic ferromagnet for
the case where the temperatures of the martensitic ($T_m$) and
magnetic ($T_C$) transitions are close. In this situation, to find
the equilibrium states of the cubic ferromagnet, the free energy
(12) must be minimized with respect to the variables $m_x$, $m_y$,
$m_z$, $e_2$, and $e_3$. To be more definite, we will assume here
that $b > 0$ and that the signs of the other quantities are the
same as in the case where $T_m < T_C$ (the case with $b < 0$ and
$K < 0$ has been examined in Ref.~[207]).

Minimization of Eqn~(12) leads to the following equilibrium states
of the ferromagnet.

1. The paramagnetic cubic (PC) phase:

\begin{equation}
m_x = m_y = m_z = 0, \quad e_2 = e_3 = 0.
\end{equation}

\noindent The PC phase is stable at $\alpha \ge 0$ and $a\ge 0$.

2. The paramagnetic tetragonal (PT) phase:

\begin{equation}
m_x = m_y = m_z = 0, e_2 = 0, e_3=-\frac{b+\sqrt{b^2-4ac}}{2c}.
\end{equation}

\noindent The PT phase is stable at

\begin{equation}
\begin{split}
 \\ a\le \frac{b^2}{4c}, \quad
a\ge\frac{b^2}{4c}-&\bigg(\frac{\sqrt{6}}{4}\frac{\alpha}{B_2}\sqrt{c}-\frac{b}{2\sqrt{c}}\bigg)^2,\\
\alpha\ge &\frac{\sqrt{6}B_2b}{3c}.
\end{split}
\end{equation}

3. The ferromagnetic cubic (FC) phase and the ferromagnetic
tetragonal (FT) phase with magnetization along the [001] axis,

\begin{equation}
m_x=m_y=0, \quad m_z^2=-\frac{\alpha+(2\sqrt{6}/3)B_2e_3}{\delta},
\end{equation}

\noindent and with strains specified by the equations

\begin{equation}
e_2=0, \quad ae_3+be^2_3+ce^3_3+\frac{\sqrt{6}B_2m^2_z}{3}=0.
\end{equation}

\noindent The FC phase is stable at

\begin{equation}
\begin{split}
\alpha\le a, \quad \alpha\ge&\frac{\sqrt{6}b}{54B_2c^2}(2\delta
b^2+12cB^2_2-9\delta ac)-\\ &-\frac{\sqrt{6}}{27B_2c^2}(\delta
b^2+4cB^2_2-3\delta ac)^{3/2},
\end{split}
\end{equation}

\noindent while the FT phase is stable at

\begin{equation}
\begin{split}
\alpha \le & \frac{\sqrt{6}b}{54B_2c^2}(2\delta
b^2+12cB^2_2-9\delta ac)+\\ +&\frac{\sqrt{6}}{27B_2c^2}(\delta
b^2+4cB^2_2-3\delta ac)^{3/2},\\
a\le&\frac{b^2}{4c}-\bigg(\frac{\sqrt{6}}{4}\frac{\alpha\sqrt{c}}{B_2}-\frac{b}{2\sqrt{c}}\bigg)^2
\quad (\alpha \ge 0).
\end{split}
\end{equation}

\noindent The stability region for the phases FC and FT is also
bounded by the inequalities

\begin{equation}
\alpha\ge-\delta, \quad a\ge-\frac{3c(\alpha+\delta)^2}{8B^2_2} +
\frac{\sqrt{6}b(\alpha+\delta)}{4B_2} +
\frac{4B^2_2}{3(\alpha+\delta)},
\end{equation}

\noindent which follow from the condition that $m_2 \le 1$.

Symmetry considerations suggest that, in addition to the above
states, other equilibrium phases with energies and stability
regions coinciding with the former ones can take place in the
crystal. Among these are tetragonal paramagnetic phases with
strains along the [100] and [010] axes, ferromagnetic cubic phases
with magnetizations along the [100] and [010] axes, and tetragonal
phases with strains and magnetizations along the [100] and [010]
axes.

\begin{figure}[t]
\includegraphics[width=7cm]{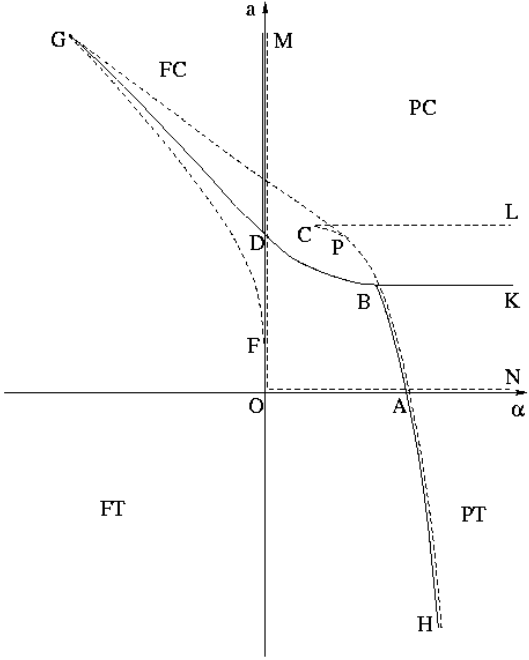}
\caption{Phase diagram of a cubic ferromagnet at $T_m \sim T_C$ in
the $(\alpha, a)$ plane. PC is the paramagnetic cubic phase, PT is
the paramagnetic tetragonal phase, FC is the ferromagnetic cubic
phase with small tetragonal distortions, and FT is the
ferromagnetic tetragonal phase. The magnetization in FS and FT is
directed along the [001] axis. The solid curves represent the
phase transition lines and the dashed curves the lines of loss of
stability of the phases.}
\end{figure}

As in the case where $T_m < T_C$, in the stability regions of the
stable phases there can also be metastable phases of other
symmetries with energies close to those of the stable phases.
Similarly, analysis of the distortions of the cubic lattice in the
FC and FT phases, which are determined by equations~(24), shows
that, from the viewpoint of symmetry, these phases have the same,
tetragonal, symmetry. The two phases differ in the values of the
spontaneous strains $e_3$. In the FC phase they are determined
chiefly by magnetostriction, while in the FT phase they are
determined chiefly by structural distortions that emerge as a
result of the transition to a martensitic state.

Figure~17 is the phase diagram of a cubic crystal at $T_m \sim
T_C$ in the $(\alpha, a)$ plane. The following phase transitions
from the cubic phase PC are possible. On the line $BK$, given by
the equation $a = b^2/4c$, there occurs a first-order structural
phase transition to a paramagnetic tetragonal phase PT with large
lattice distortions (a martensitic transformation). Along the line
$DM$ ($\alpha = 0$) there proceeds a second-order magnetic phase
transition to a ferromagnetic cubic phase FC with small tetragonal
lattice distortions. On the line $DB$ there occurs a first-order
coupled structural and magnetic phase transition to a
ferromagnetic tetragonal phase FT with large tetragonal lattice
distortions. The equation of this line can be found by equalizing
the energies of the phases PC and PT. In addition to the above
PT--PC transition on the line $BK$, there can occur a second-order
isostructural magnetic phase transition from the paramagnetic
tetragonal phase PT to a ferromagnetic tetragonal phase FT along
the line $BH$. The equation of this line follows from the second
condition of stability of the FT phase in Eqn~(26) if the
inequality sign is replaced by the equality sign. Between the
ferromagnetic phases FC and FT there can occur a first-order
isostructural phase transition on the line $GD$. The equation of
this line is

\begin{equation}
a=\frac{2}{9}\frac{b^2}{c}+\frac{4}{3}\frac{B^2_2}{\delta}-
\frac{\sqrt{6}B_2c\alpha}{6\delta}.
\end{equation}

\noindent The given transition is accompanied by a discontinuity
in the strain $e_3$ and is a martensitic transformation. Its
termination point in the ($\alpha, a$) plane may be the point $G$.
This happens if the point $G$ is located to the right of the line
of stability of the FC and FT phases in Eqn~[27] (we have not
depicted this line so as not to clutter up the figure). Analysis
of inequality~(27) and equation~(28) combined with conditions~(25)
and (26) of stability of the FC and FT phases shows that the point
of termination of the phase transition of FT and FC phases will
exist only at large values of the magnetoelastic constant $B_2
\sim b^3/c^2$. In this case the transition between the FT and FC
phases to the left of point $G$ proceeds smoothly, without a
discontinuity in the strain $e_3$.

The region of absolute stability of the PC phase is limited by the
lines $OM$ and $ON$. For the PT phase this region is limited by
the lines $LC$ and $CH$. The FT phase is absolutely stable in the
region to the left of the line $GH$ and the FC phase above the
curve $GFM$. The points $D$ and $B$ are critical: at these points
the line of second-order phase transitions splits into two lines
of first-order phase transitions. The coordinates of points $D$
and $B$ are

\begin{equation*}
\bigg(\frac{2b^2}{9c}+\frac{4B^2_2}{3\delta},\;\; 0\bigg), \qquad
\bigg(\frac{2b^2}{9c},\;\; \frac{4\sqrt{6}B_2b}{9c}\bigg).
\end{equation*}

Our analysis of the effect of magnetoelastic interaction on the
phase diagrams of the cubic ferromagnet shows that when the first
anisotropy constant $K$ is positive, structural phase transitions
are not accompanied by reorientation of magnetization. The
explanation is that, strictly speaking, the magnetoelastic
interaction already in the cubic phase lowers the symmetry of this
phase to tetragonal. The symmetry of the low-temperature phase
proves to be either tetragonal ($b > 0$) or orthorhombic ($b <
0$). And because the high- and low- temperature phases contain the
same symmetry elements (say, 2-fold and 4-fold axes), no
reorientation of magnetization occurs in structural transitions.

When the symmetries of the high- and low-temperature phases
coincide perfectly ($b > 0$), the line of structural phase
transition may have a termination point. To the right of this
point a structural transition is accompanied by a discontinuity of
strains and by hysteresis and represents a martensitic
transformation. To the left of this point there is no transition
the strains gradually change their nature from quasicubic to
tetragonal, and there is no hysteresis. When $b < 0$, the
symmetries of the low- and high-temperature phases do not
coincide. In this case, at large values of $|b|$ the structural
transition between the phases is a first-order phase transition (a
martensitic transformation), while at small values of $|b|$ the
transition is of the second order. Thus, at $b < 0$ we have a
critical point in the phase diagram (at this point the type of
transition changes). The coordinates of the structural-transition
termination point at $b > 0$ and the critical point at $b < 0$ are
determined by the magnetostriction constant $B_2$. The presence of
magnetoelastic interaction leads to a situation in which within a
certain interval of the parameters of the cubic ferromagnet there
occur first-order coupled structural and magnetic phase
transitions. The size of this interval in the phase diagram where
such transitions occur is determined by the magnitude of the
magnetoelastic interaction.

\subsection{\textit{T vs. x} phase diagram at $T_m \approx T_C$}

The experimental data (e.g., see Refs~[66, 67, 91]) suggest that
in Ni$_{2+x}$Mn$_{1-x}$Ga alloys the Curie temperature and the
martensitic transformation temperature vary with concentration $x$
almost linearly. This fact can be used to build a theoretical $T
vs. x$ phase diagram of the given alloys. To this end we write the
coefficients $a$ and $\alpha$ in the free energy (12) in the form

\begin{equation}
a=\frac{a_0(T-T_m)}{T_m}, \quad
\alpha=\frac{\alpha_0(T-T_C)}{T_C},
\end{equation}

\noindent where $T_m = T_{m0} + \sigma x$ and $T_C = T_{C0} -
\gamma x$, with $T_{m0}$ and $T_{C0}$ the temperatures of the
martensitic and magnetic transitions at $x = 0$, and $\sigma$ and
$\gamma$ are proportionality factors. The values of $T_{m0}$,
$T_{C0}$, $\sigma$, and $\gamma$ can be found from experiments.

\begin{figure}[t]
\includegraphics[width=7cm]{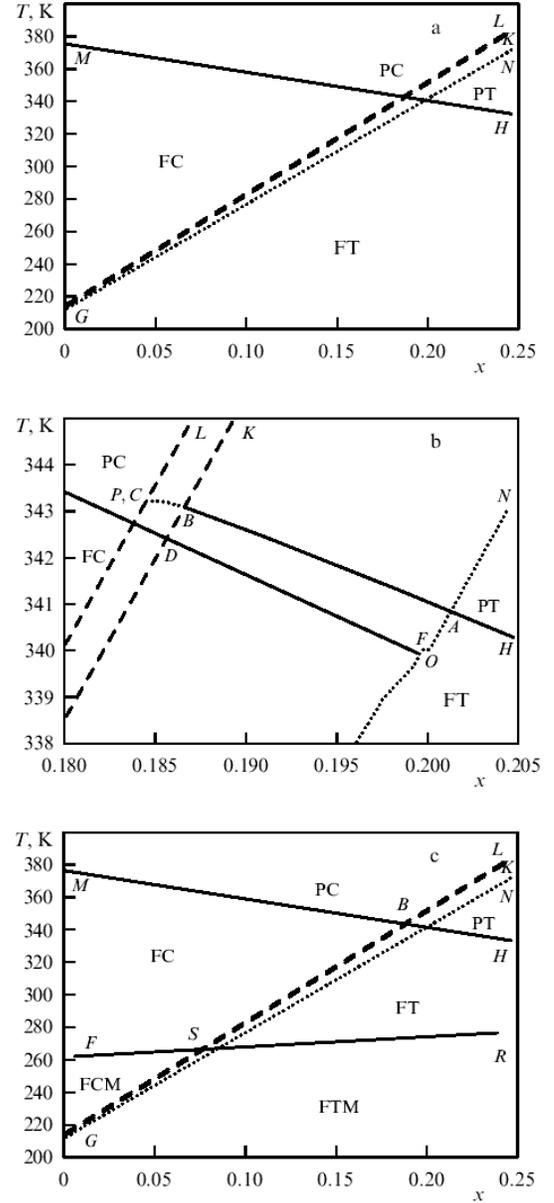}
\caption{Theoretical $T$ vs. $x$ phase diagram of cubic
ferromagnetic alloys Ni$_{2+x}$Mn$_{1-x}$Ga without modulation of
the crystal lattice (a) [(b) shows the region where the
temperatures $T_m$ and $T_C$ intersect on a larger scale], and
with modulation of the crystal lattice (c).}
\end{figure}

Figure~18a shows a phase diagram of Ni$_{2+x}$Mn$_{1-x}$Ga alloys
built numerically with allowance for Eqn~(29). The following
values of the parameters in Eqns~(12) and (29) (see Refs [14, 20,
21, 23, 66, 67, 80, 91, 123, 223]) were used in calculating this
diagram:

\begin{equation}
\begin{split}
a_0=10^{11}~\mathrm{erg/cm}^{-3}, \;
\alpha_0=-10^9~\mathrm{erg/cm}^{-3}, \\ T_{m0}=202~\mathrm{K}, \;
T_{C0}=375~\mathrm{K}, \; \sigma = 700~\mathrm{K}, \\ \gamma =
175~\mathrm{K}, \; b=3\times10^{11}~\mathrm{erg/cm}^{-3},\\
c=3\times10^{12}~\mathrm{erg/cm}^{-3}, \;
B_2=1.5\times10^7~\mathrm{erg/cm}^{-3},\\
K=4\times10^4~\mathrm{erg/cm}^{-3}, \; \delta =
10^9~\mathrm{erg/cm}^{-3}.
\end{split}
\end{equation}

In view of the smallness of the magnetoelastic constant $B_2$, the
region in Fig.~18a where the temperatures $T_m$ and $T_C$
intersect is not resolved on the given scale. To resolve it, in
Fig.~18b we show this region on a larger scale. Figures~18a and
18b suggest that with the parameters chosen according to Eqn~(30),
the region $DB$ where first-order structural and magnetic phase
transitions coexist is small and is located near $x \approx
0.187$. The size of this interval is determined by the
magnetoelastic constant $B_2$. The phase diagram in Figs~18a and
18b is in good agreement with the experimental $T vs. x$ diagram
[66, 67, 91] (see Fig.~10). We also note that in the region where
a coupled structural and magnetic phase transition exists (line
$DB$ in Fig.~18b), the Curie temperature increases with
concentration. Outside this region the Curie temperature always
decreases (with $x$ increasing or decreasing). This fact also
agrees with the experimental data (see Section~5.1).

As noted earlier [see Eqn~(18)], within the region of existence of
stable phases in which the lattice parameter ratio $c/a$ is less
than unity, there can also be metastable phases (with energies
close to those of the stable phases), in which the lattice
parameter ratio $c/a$ is greater than unity. Figures~18a and 18b
show that within a narrow interval of concentrations $x \approx
0.19$ near the point of intersection of $T_m$ and $T_C$, phases
with cubic symmetry may exist in addition to martensitic phases of
different symmetries and different lattice parameter ratios $c/a$.
Crystal phases of different symmetries and different values of
$c/a$ have indeed been observed in experiments [99, 221, 222] in
alloys with $x \approx 0.19$.

\subsection{\textit{T vs. x} phase diagram at $T_m \approx T_C$ with allowance
for the modulation order parameter}

To find all the equilibrium states of the Ni$_{2+x}$Mn$_{1-x}$Ga
alloys when the modulation order parameter is taken into account,
we must minimize expression (12) in the order parameters $e_2$,
$e_3$, $|\psi|$, $m_x$, $m_y$, and $m_z$. The resulting system of
nonlinear algebraic equations can be solved only numerically. To
do the necessary numerical calculations, the following values of
parameters in equation (12) were chosen on the basis of the
existing experimental data [14, 20, 21, 23, 66, 67, 80, 91, 123,
223] (in units of erg/cm$^{-3}$):

\begin{equation*}
\begin{split}
D_2=10^3, \quad A_0=10^{23}, \quad B^{\prime}=10^{38}, \quad
C^{\prime}=10^{55}, \\ N^{\prime}_1=10^3, \quad N_2=-100, \quad
N^{\prime}_3=-100.
\end{split}
\end{equation*}

\noindent The other parameters were taken from Eqn~(30). The
experimental data (see Fig.~10) suggest that the premartensitic
transformation temperature $T_P$ is practically independent of the
composition of the alloys investigated. This makes it possible to
write the parameter $A$ in Eqn~(12) in the form $A = A_0(T -
T_{P0})/T_{P0}$, where $T_{P0} = 260$~K.

The \textit{T vs. x} phase diagram of Heusler alloys
Ni$_{2+x}$Mn$_{1-x}$Ga calculated with the given values of the
parameters is shown in Fig.~18c. Clearly, with the parameters
chosen in this way the following phases emerge as a result of
concentration and temperature variations: the paramagnetic cubic
(PC), the paramagnetic tetragonal (PT), the ferromagnetic
quasicubic (FC), the ferromagnetic tetragonal (FT), the
ferromagnetic quasicubic modulated (FCM), and the ferromagnetic
tetragonal modulated (FTM). In all the ferromagnetic phases
magnetization is directed along the [001] axis. $MB$ and $BH$ are
the lines of second-order magnetic phase transitions between the
paramagnetic and ferromagnetic cubic and tetragonal phases,
respectively. $FS$ is the line of the first-order phase transition
between the ferromagnetic phases FC and FCM. $SR$ is the line of a
similar transition between the tetragonal phases FT and FTM. These
first-order phase transitions are accompanied by the emergence of
lattice modulation. The hysteresis of the given transitions with
the selected values of parameters is very small and therefore is
not resolved on the scale of Fig.~18c. $GS$, $SB$, and $BK$ are
the lines of the martensitic phase transitions FCM--FTM, FC--FT,
and PC--PT, respectively. A characteristic feature of these
transitions is that they are accompanied by the emergence of large
tetragonal strains of the lattice. $GL$ and $GN$ are lines of loss
of stability of the phases FTM, FT, PT and FCM, FC, PC,
respectively. Figure~18c shows that, compared to a transition to a
modulated state, the martensitic transition is accompanied by a
large hysteresis. This fact is in good agreement with the
experimental data from resistivity measurements (see Fig.~11). In
the concentration interval $x \approx 0.06-0.1$, the hysteresis
regions of the premartensitic and martensitic transitions overlap,
with the result that these transitions are partially superimposed
and are poorly resolved in experiments.

\subsection{Effect of magnetic field on martensitic structure}

Analysis of the phase diagrams done in the previous sections shows
that at $T < T_m$ the Heusler alloys Ni$_{2+x}$Mn$_{1-x}$Ga divide
into structural domains. Defects in the crystal lattice play an
important role in the nucleation of martensitic structures
[224-226]. Detailed analysis of martensitic domain structures has
been carried out in Refs [226-228]. Curnoe and Jacobs [229]
proposed a model of a twin boundary separating two tetragonal
variants. The structural domains in the ferromagnetic alloys
Ni$_{2+x}$Mn$_{1-x}$Ga, in turn, divide into magnetic domains.
When an external magnetic field is applied, both the magnetic and
the structural domains undergo a transformation [75, 154,
230-238].

L'vov \textit{et al.} [231] calculated the temperature curves for
the magnetization of a ferromagnetic martensite at different
values of the external magnetic field strength. They found that
the presence of a martensitic domain structure helps to explain
the temperature dependences of magnetization observed in
experiments (see Fig. 4).

James and Wuttig [154] found the field dependence of the strains
induced in an alloy by numerical minimization of the alloy energy
averaged over the different martensitic variants. They assumed
that each martensitic variant that is possible for the specific
sample enters into the total magnetization and strain with its own
volume fraction. The total energy was expressed by the sum of
Zeeman energy, the external stresses energy and the
demagnetization energy, which, in turn, were expressed in terms of
the averaged magnetization and strain. Numerical minimization of
the total energy made it possible to determine the volume
fractions of the optimal martensitic variants for the given value
of the magnetic field, which made it possible to construct the
field dependencies of the strains induced by the magnetic field in
the samples. A fuller analysis of the effect of an external
magnetic field on the martensitic domain structure of
Ni$_{2+x}$Mn$_{1-x}$Ga alloys was carried by O'Handley~[230], who,
however, did not examine the specific magnetic domain structure
observed in experiments.

The structural and magnetic domains in the martensitic phase of
ferromagnetic Heusler alloys Ni$_{2+x}$Mn$_{1-x}$Ga have been
observed in experiments by Chopra \textit{et al.}~[121] (who used
the photomicrography method) and by Pan and James~[85] (who used
the magnetic-force microscopy method). On the basis of the
experimental results [85, 121] and the theoretical analysis of the
phase diagram of Ni$_{2+x}$Mn$_{1-x}$Ga alloys [66, 67, 91,
200-215] Buchel'nikov \textit{et al.} [236-238] proposed a model
of the self-consistent structure of martensitic and magnetic
domains, studied the behavior of this structure under an applied
magnetic field, and calculated the strains induced by the magnetic
field in the Ni$_{2+x}$Mn$_{1-x}$Ga alloys.

As a result of the transition from the high-temperature cubic
modification to the low-temperature tetragonal modification, the
Ni$_{2+x}$Mn$_{1-x}$Ga single crystal divides into three types of
martensitic domains, each of which corresponds to a strain
(compression or extension) of the crystal lattice along directions
of the \{100\} type. In the ferromagnetic crystal, the structural
domains, in turn, divide into magnetic domains. The direction of
the magnetization in each magnetic domain coincides with the
principal crystallographic axis of the structural domain. For
definiteness, in the structural domain of the first type the
tetragonal axis is directed along the \textbf{x} axis, while in
the domains of the second and third types this axis is directed
along the \textbf{y} and \textbf{z} axes, respectively. It is
assumed that each structural domain divides into 180-degree
magnetic domains.

When there is no magnetic field, all structural domains have the
same energy, with the result that the volumes occupied by each
type of domain are the same, too. When a magnetic field directed
along the \textbf{x} axis is switched on, the volume of domains of
the first type increases at the expense of the domains of the
second and third types. In the presence of the field the volume
fractions of domains of different types can be written as follows:

\begin{equation}
f_1 = \frac{1}{3}+\xi, \quad f_{23}=\frac{2}{3}-\xi,
\end{equation}

\noindent where $f_1$ is the volume fraction of domains of the
first type, $f_{23}$ is the volume fraction of domains of the
second and third types, and $\xi$ is a parameter that takes into
account the variation of domain volumes under the applied magnetic
field.

The results obtained in the previous sections suggest that in a
zero magnetic field the cubic ferromagnet can transform into
tetragonal phases with magnetizations along axes of the \{001\}
type. For instance, in the phase with $m_z = 1$ $(\mathbf{M}
||[001])$ the spontaneous strain tensor has the form

\begin{gather}
e_{ij} = \begin{Vmatrix}e_{xx} & 0 & 0 \\ 0 & e_{xx} & 0 \\ 0 & 0
& e_{zz} \end{Vmatrix}
\end{gather}

\noindent where

\begin{equation*}
\begin{split}
e_{xx}=-\frac{B_1}{3(C_{11}+2C_{12}})-\frac{e_0}{\sqrt{6}},\\
e_{zz}=-\frac{B_1}{3(C_{11}+2C_{12}})+\frac{2e_0}{\sqrt{6}},
\end{split}
\end{equation*}

\noindent and the constant $e_0$ can be found by solving the
equation

\begin{equation*}
ae_0+be^2_0+ce^3_0+\frac{2B_2}{\sqrt{6}}=0.
\end{equation*}

\noindent For the other two phases with magnetizations along the
[100] and [010] axes, the spontaneous strain tensor can be
obtained from Eqn~(32) through a cyclic permutation of the
indices. Thus, in the ferromagnet there can exist three tetragonal
phases with the same energy, with the result that there can be
three types of structural domains in the crystal. The first terms
in the expressions for $e_{ik}$ describe the changes in strains
caused by bulk magnetostriction. Compared to effects associated
with structural lattice distortions, these terms are small and can
be ignored.

Plugging Eqn~(32) into the total energy (12) of the cubic
ferromagnet leads to the following expression for the energy of
each type of domains in the system of coordinates associated with
the crystal lattice:

\begin{equation}
\begin{split}
F&= \frac{1}{2}ae^2_0+\frac{1}{3}be^3_0+\frac{1}{4}ce^4_0+\\
&+Km^2_z+K_1(m^2_xm^2_y+m^2_ym^2_z+m^2_zm^2_x)-\mathbf{HM},
\end{split}
\end{equation}

\noindent where $K = (\sqrt{6}/2)B_2e_0$ is the induced uniaxial
anisotropy and \textbf{H} is the internal magnetic field. Tickle
and James~[80] measured the magnetic anisotropy of
Ni$_{2+x}$Mn$_{1-x}$Ga alloys and found that the uniaxial
anisotropy is much larger than the cubic one, so that in all
further calculations the cubic anisotropy is dropped.

We write the energy of the domain structure in the following form:

\begin{equation}
F=F_c+F_m+F_Z
\end{equation}

\noindent where $F_e$ is the energy of the elastic subsystem,
$F_m$ is the energy of the magnetic subsystem, and $F_Z$ is the
Zeeman energy.

Next we write the energy of the elastic subsystem as follows:

\begin{equation}
\begin{split}
F_e&=\frac{1}{2}\tilde{C}\big(\langle e_{xx}\rangle^2+\langle
e_{yy}\rangle^2+\langle e_{zz}\rangle^2\big)+\\
&+\tilde{C}_{12}\big(\langle e_{xx}\rangle\langle
e_{yy}\rangle+\langle e_{xx}\rangle\langle e_{zz}\rangle+\langle
e_{zz}\rangle\langle e_{yy}\rangle\big),
\end{split}
\end{equation}

\noindent where $\langle\dots\rangle$ stands for volume averaging,
and $\tilde{C}_{ijkl}$ is the effective elastic modulus tensor for
the Ni$_{2+x}$Mn$_{1-x}$Ga alloy with the domain structure in
question. According to Eqn~(31), the average values of the
components of the strain tensor of the alloy can be written as

\begin{equation}
\langle e_{ij}\rangle=\bigg(\frac{1}{3}+\xi\bigg)e^{(1)}_{ij} +
\bigg(\frac{2}{3}-\xi\bigg)
\bigg(\frac{1}{2}e^{(2)}_{ij}+\frac{1}{2}e^{(3)}_{ij}\bigg),
\end{equation}

\noindent where $e^{(k)}_{ij}$ is the strain tensor for $k$-type
domains defined in Eqn~(32). Plugging these strains into Eqn~(36)
yields the following formula for the strain tensor averaged over
the sample's volume:

\begin{gather}
\langle e_{ij}\rangle = \frac{\sqrt{6}}{4} \begin{pmatrix}2e_3\xi
& 0 & 0 \\ 0 & -e_3\xi & 0 \\ 0 & 0 & -e_3\xi \end{pmatrix}.
\end{gather}

Plugging Eqn~(37) into expression (35), we arrive at the following
expression for the elastic energy of the domain structure in
question:

\begin{equation}
F_c = \frac{9\tilde{C}e^2_3\xi^2}{8},
\end{equation}

\noindent where $\tilde{C} = \tilde{C}_{11} - \tilde{C}_{12}$.

According to our choice of the direction of the external magnetic
field, magnetization in domains of the first type is directed
parallel or antiparallel to the field, while in domains of the
second and third types it is perpendicular to the field. In view
of this, in domains of the first type under magnetization there
can only be processes in which the domain walls are displaced
while in domains of the second and third types there can only be
processes in which magnetization is rotated. In this case the
anisotropy energy and the Zeeman energy can be written as follows:

\begin{equation}
F_m = \bigg(\frac{2}{3}-\xi\bigg)K\cos^2\varphi, \quad F_Z =
-H\langle M_x\rangle.
\end{equation}

\noindent Here the projection of the magnetization vector on the
direction of the magnetic field has the form

\begin{equation}
\langle M_x\rangle = \bigg(\frac{1}{3}+\xi\bigg)M^{(1)} +
\bigg(\frac{2}{3}-\xi\bigg)M_0\cos\varphi,
\end{equation}

\noindent and the internal magnetic field is given by the formula
$H = H_0-4\pi N\langle M_x\rangle$, where $H_0$ is the external
magnetic field and $N$ is the demagnetization factor. In equations
(39) and (40), $\varphi$ is the angle between the directions of
the magnetization and the magnetic field vectors in domains of the
second and third types, and $M^{(1)}$ is the average magnetization
in domains of the first type. This magnetization can be expressed
by the following formula:

\begin{equation}
M^{(1)} =
\begin{cases}
\frac{HM_0}{H_C}, & H < H_C,\\ M_0, & H > H_C,
\end{cases}
\end{equation}

\noindent where the phenomenological parameter $H_C$ is the field
strength at which all processes of displacement of magnetic domain
walls in martensitic domains of the first type are terminated.

Plugging Eqns~(38) and (39) into formula~(34), we arrive at the
following expression for the energy of the domain structure:

\begin{widetext}

\begin{equation}
F=
\frac{9\tilde{C}e^2_3\xi^2}{8}+\bigg(\frac{2}{3}-\xi\bigg)K\cos^2\varphi-
H\bigg[\bigg(\frac{1}{3}+\xi\bigg)M^{(1)}+\bigg(\frac{2}{3}-\xi\bigg)M_0\cos\varphi\bigg].
\end{equation}

The equilibrium values of the parameters $\xi$ and $\varphi$ of
the domain structure in question can be found by minimizing the
expression (42) with respect to these variables. This leads to the
following equations for finding $\xi$ and $\varphi$:

\begin{equation*}
32\pi NKM^2_0\cos^3\varphi-96\pi NKM_0M^{(1)}\cos^2\varphi+
\big[6\tilde{C}e^2_0(8\pi NM^2_0+3K)+64\pi
NK(M^{(1)})^2\big]\cos\varphi-3\tilde{C}e^2_0M_0(3H_0-8\pi
NM^{(1)})=0,
\end{equation*}
\begin{equation}
\xi=-\frac{4}{3}\frac{(M^{(1)}-M_0\cos\varphi)(16\pi
NM_0\cos\varphi-3H_0+8\pi
NM^{(1)})-3K\cos^2\varphi}{9\tilde{C}e^2_0+32\pi
N(M^{(1)}-M_0\cos\varphi)^2}.
\end{equation}

\end{widetext}

Plugging the solution of equations (43) into the expressions for
the strain tensor (37) and the magnetization (40), we can find the
dependence of these characteristics of Ni$_{2+x}$Mn$_{1-x}$Ga
alloy on the size of the external magnetic field. The results of
numerical calculations of the field dependencies of the
magnetization, the parameter $\xi$, and the strain $\langle
e_{xx}\rangle$ for different values of the effective elastic
modulus $\tilde{C}$ and the demagnetization factor $N$ are shown
in Fig.~19. The following values of the parameters of
Ni$_{2+x}$Mn$_{1-x}$Ga alloy [80, 85] were used in the
calculations:

\begin{equation*}
\begin{split}
x&\approx0,  \qquad M_0=500~\mathrm{G}, \qquad e_0=0.06, \\
K&=2\times10^6~\mathrm{erg/cm}^{-3}, \;\; N=0.02, \;\;
H_C=2000~\mathrm{Oe}.
\end{split}
\end{equation*}

\noindent These calculations also took into account the fact that
when the parameter $\xi$ reaches the value of 2/3, the alloy
becomes a single-domain alloy, the strain $\langle e_{xx}\rangle$
reaches the value of roughly 0.06 (this is determined by the ratio
of the crystal lattice parameters in the tetragonal phase, $c/a =
0.94$), and no further strain of the alloy under a magnetic field
occurs.

\begin{figure}[b]
\includegraphics[width=7cm]{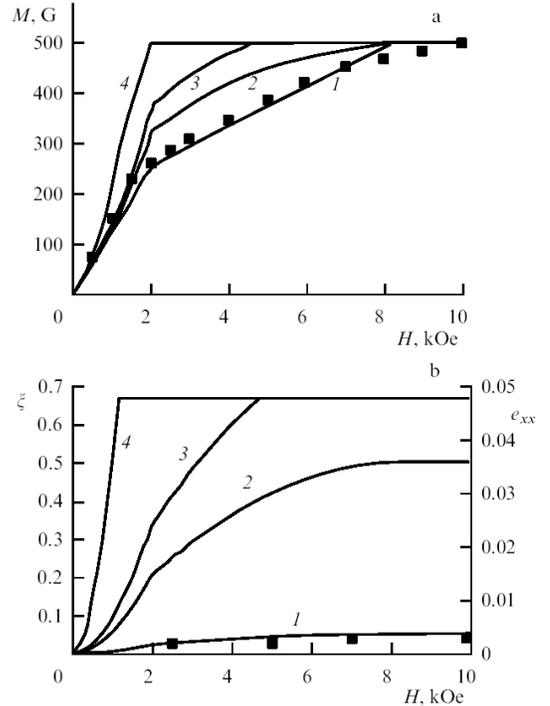}
\caption{Field dependencies of the magnetization $M$ (a), and the
volume fraction $\xi$ of the favorably located martensitic domains
and of the induced strains $e_{xx}$ (b) for different values of
the effective elastic modulus $\tilde{C}$ (in units of
erg/cm$^{-3}$): $1 - 5\times10^9$, $2-5\times10^8$,
$3-3\times10^8$, and $4- 5\times10^7$; the black squares represent
the experimental data of Wu \textit{et al.} [123].}
\end{figure}

The field dependencies of $\langle M_x\rangle$, $\xi$, and
$\langle e_{xx}\rangle$ have kinks at field strengths $H_0 \sim
H_C$ and $H_0 \sim H_1 = H_A + 8\pi N\langle M_x\rangle$, where
$H_A = 2K/M_0$. The explanation is as follows. When $H_0 < H_C$,
the increase in magnetization is caused by the displacement of the
magnetic domain walls in structural domains of the first type and
by rotation of magnetization vectors in structural domains of the
second and third types. At $H_0 = H_C$ the magnetic domains in the
structural domains of the first type disappear completely, and the
further increase in magnetization in fields $H_C < H_0 < H_1$ is
due solely to the rotation of magnetization in the structural
domains of the second and third types. At $H_0 = H_1$ the vectors
\textbf{M} in the structural domains of the second and third types
align themselves with the direction of the external field, so that
the sample's magnetization reaches its saturation value.

Figure~19 suggests that the transformation of magnetic domains
under an external field takes place simultaneously with the
transformation of the structural domains. As the magnetic field
gets stronger, the structural domains of the first type grow
bigger. In fields higher than $H_1$, when the magnetization is the
same in all structural domains, the Zeeman energy of these domains
is also the same. Since it is due to the increase in the Zeeman
energy that structural domain walls begin to move under a magnetic
field, at $H_0 > H_1$ a further growth of the structural domains
of the first type at the expense of the domains of the second and
third types becomes impossible.

The limit value of the volume fraction that can be reached in
magnetization saturation is given by the formula

\begin{equation}
\xi_{max}=\frac{2M_0H_A}{9\tilde{C}e^2_0},
\end{equation}

\noindent and the corresponding value of the strain $\langle
e_{xx}\rangle$ caused by the magnetic field is given by the
formula

\begin{equation}
\langle
e_{xx}\rangle_{max}=\frac{\sqrt{6}M_0H_A}{9\tilde{C}e^2_0}.
\end{equation}

Figure 19 clearly shows that at large values of the effective
elastic modulus $\tilde{C}$ the values of $\xi_{max}$ and $\langle
e_{xx}\rangle_{max}$ are smaller than the maximum values
determined by changes in the lattice. From equations (44) and (45)
it follows that to achieve maximum strains induced by a magnetic
field in Ni$_{2+x}$Mn$_{1-x}$Ga alloys, the elastic energy must be
of the order of the anisotropy energy. This can be achieved at
small values of the elastic modulus $\tilde{C}$ near the
martensitic transformation point. The results of the calculated
field dependencies of magnetization and strain are in good
agreement with the experimental data [80, 19, 123].

\subsection{Shift in the structural phase transition temperature
caused by a magnetic field}

The shift in the temperature of a martensitic transformation
caused by a magnetic field $H$ and by mechanical stresses $P$ has
been studied in Refs [108-111, 198, 206, 239-245]. The $T$ vs. $P$
and $T$ vs. $H$ phase diagrams of some Ni$_{2+x}$Mn$_{1-x}$Ga
alloys have been calculated in Refs [198, 206, 239, 240, 242-245].

There are two reasons why a magnetic field shifts the martensitic
transition temperature. The first is the displacement of the phase
transition line caused by changes in the free energies (12) of the
austenitic and martensitic phases in the presence of a magnetic
field. The second reason is the change in the thermodynamic
equilibrium temperature of the austenitic and martensitic phases
in a magnetic field. The first factor changes the phase transition
temperature only slightly but considerably narrows the hysteresis
of the martensitic transition in relatively weak magnetic fields
[202, 206, 239]. This happens because the energy of the
magnetoelastic interaction is low compared to the Zeeman energy
already in fields of several kOe. The second factor can
significantly change the phase transition temperature. It is this
effect that we will now consider.

Experimenters usually measure the temperatures of the beginning of
the martensitic ($T_{ma}$) and austenitic ($T_{am}$)
transformations. Their dependence on the magnetic field strength
can be found from the phase equilibrium condition (the Clausius--
Clapeyron equation) [109]. At $T = T_{am}$ the phase equilibrium
condition can be written as

\begin{widetext}

\begin{equation}
\Phi_m(T_{am},H)-\Phi_a(T_{am},H)=
\Delta\phi(T_{am})-(M_mV_m-M_aV_a)H+\phi^{am}_c+\phi^{am}_s=0.
\end{equation}

\noindent at $T = T_{ma}$ this condition is

\begin{equation}
\Phi_a(T_{am},H)-\Phi_m(T_{ma},H) =
\Delta\phi(T_{ma})-(M_aV_a-M_mV_m)H+\phi^{ma}_c+\phi^{ma}_s=0.
\end{equation}

\end{widetext}

\noindent Here, $\Phi_{a,m}$ are the thermodynamic potentials of
the austenitic and martensitic phases, $M_{a,m}$ and $V_{a,m}$ are
the magnetizations and the volumes of these phases, $\phi_{c,s}$
are additional energies related to the emergence of additional
stresses due to phases matching at the austenite--martensite twin
boundaries and to the surface energy of these twin boundaries, and
$\Delta\phi(Y) = Q(T-T_m)/T_m$, with $Q$ and $T_m$ are,
respectively, the latent heat and the temperature of the phase
transformation.

Combining equations (46) and (47), we obtain the following formula
for the field dependence of the temperatures $T_{ma}$ and
$T_{am}$:

\begin{equation}
\begin{split}
&T_{am,ma}=\\
&=T_m\bigg\{\frac{1+\big[(M_mV_m-M_aV_a)H\mp(\phi^{am,ma}_c+\phi^{am,ma}_s)\big]}{Q}\bigg\}.
\end{split}
\end{equation}

To find the unknown quantities $M_{a,m}$ and $T_m$ in Eqn~(48),
Cherechukin \textit{et al.}~[109] used the results of minimization
of the free energy (12) with allowance for the Zeeman term
$-M_ZH$. Figure~20 shows the field dependencies of temperatures
$T_{ma}$ and $T_{am}$ calculated by using Eqn~(48) with the
numerical values of the parameters in the expression for the free
energy (12) adopted by Cherechukin \textit{et al.} [109]. Here we
also sow the experimental results for a
Ni$_{2.15}$Mn$_{0.81}$Fe$_{0.04}$Ga alloy. Clearly, there is good
agreement between theory and experiment.

\begin{figure}[b]
\includegraphics[width=7cm]{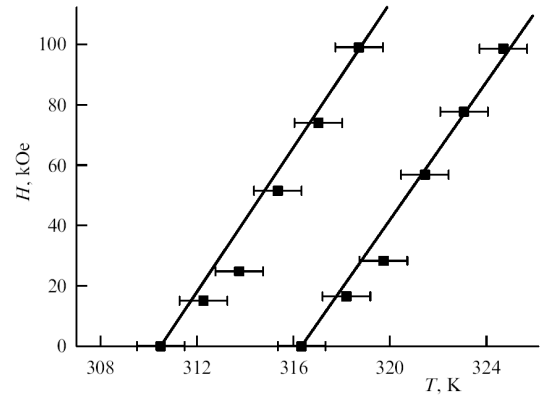}
\caption{Experimental and theoretical dependencies $T_{am}(H)$ and
$T_{ma}(H)$ for the Ni$_{2.15}$Mn$_{0.81}$Fe$_{0.04}$Ga alloy
[109]. The solid lines represent the theoretical results and the
black squares the experimental data.}
\end{figure}

\section{Conclusion}

Extensive studies of shape memory ferromagnets has led to the
discovery of a new mechanism of controlling the size and shape of
a solid by subjecting the samples to a magnetic field. The strains
that develop as result of such action exceed the record values of
magnetostrains in magnetostrictors. There are already certain
advances in the practical applications of shape memory
ferromagnets (see, e.g., Ref. [246]). The introduction of
actuators and sensors whose operation is based on this class of
materials requires, however, solving a whole range of problems.
The main factors here are the increase of durability of the
materials, the reproducibility of reversible magnetically induced
strains, the study of the role of aging, etc. An important area of
research of Ni$_{2+x+y}$Mn$_{1-x}$Ga$_{1-y}$ alloys is the study
of ways of improving the dynamic characteristics of actuators and
sensors whose operation is based on giant magnetically induced
strains [140].

In conclusion we would like to note that, notwithstanding the
enormous body of information that has already been gathered, some
of the physical properties of Heusler alloys
Ni$_{2+x+y}$Mn$_{1-x}$Ga$_{1-y}$ require further investigation.
This is true, first and foremost, of the phase diagram of the
given ternary compound, the information about which is fragmentary
[15, 141, 247, 248]. It would be especially interesting to
establish the main laws governing the concentration dependence of
magnetically induced strains in Ni$_{2+x+y}$Mn$_{1-x}$Ga$_{1-y}$
alloys. Since many characteristics of
Ni$_{2+x+y}$Mn$_{1-x}$Ga$_{1-y}$ alloys depend on composition, it
is imperative that the concentration dependencies of the
anisotropy constant, elastic moduli, and other characteristics be
measured. In the area of theory, further development of the
statistical model of phase transformations in
Ni$_{2+x+y}$Mn$_{1-x}$Ga$_{1-y}$ alloys is necessary so as to make
it possible to model the specific temporal behavior of samples
(the kinetics of the transformation) subjected to external
stresses and magnetic fields [249, 250].

\end{document}